\documentclass[
reprint,
superscriptaddress,
amsmath,
amssymb,
aip,
longbibliography,
jcp,
showpacs,
floatfix
]{revtex4-2}

\usepackage{graphicx}% Include figure files
\usepackage{dcolumn}% Align table columns on decimal point
\usepackage{bm}% bold math
\usepackage[utf8]{inputenc}
\usepackage[T1]{fontenc}
\usepackage{orcidlink}
\usepackage{mathptmx}
\usepackage{xcolor}
\usepackage{etoolbox}
\usepackage[inkscapelatex=false]{svg}
\usepackage{soul}
\usepackage{mdframed} 

\DeclareMathAlphabet{\mathcal}{OMS}{cmsy}{m}{n}
\usepackage[skip=5pt plus1pt, indent=10pt]{parskip}

\makeatletter
\def\@email#1#2{%
 \endgroup
 \patchcmd{\titleblock@produce}
  {\frontmatter@RRAPformat}
  {\frontmatter@RRAPformat{\produce@RRAP{*#1\href{mailto:#2}{#2}}}\frontmatter@RRAPformat}
  {}{}
}%
\makeatother

\def\bra#1{\mathinner{\langle{#1}|}}
\def\ket#1{\mathinner{|{#1}\rangle}}
\newcommand{\braket}[1]{ \langle #1 \rangle}
\newcommand{\ketbra}[2]{|{#1}\rangle\!\langle{#2}|}
\begin{document}

\title{Mind the Gap: From Resolving Theoretical Foundations of Chiral(ity)-Induced Spin Selectivity to Pioneering Implementations in Quantum Sensing}
\author{Yan Xi Foo\orcidlink{0009-0001-0811-4673}}
\affiliation{Quantum Science and Engineering Centre (QSec),
Nanyang Technological University, SG 639798, Singapore}
\affiliation{Quantum Biology Tech (QuBiT) Lab, Department of Electrical and Computer Engineering, University of California, Los Angeles, CA 90095, USA}
\author{Aisha Kermiche\orcidlink{0009-0002-9952-923X}}
\affiliation{Quantum Biology Tech (QuBiT) Lab, Department of Electrical and Computer Engineering,  University of California, Los Angeles, CA 90095, USA}
\author{Farhan T.\ Chowdhury\,\orcidlink{0000-0001-8229-2374}}
\affiliation{Department of Physics, University of Exeter, Stocker Road, Exeter, Devon, EX4 4QL, United Kingdom.}
\affiliation{Living Systems Institute, University of Exeter, Stocker Road, Exeter, Devon, EX4 4QD, United Kingdom.}
\affiliation{Quantum Biology Tech (QuBiT) Lab, Department of Electrical and Computer Engineering,  University of California, Los Angeles, CA 90095, USA}
\author{Clarice D. Aiello\,\orcidlink{0000-0001-7150-8387}}
\affiliation{Quantum Biology Institute, Los Angeles, CA 90095, USA}
\author{Luke D. Smith\orcidlink{0000-0002-6255-2252}}
\affiliation{Department of Physics, University of Exeter, Stocker Road, Exeter, Devon, EX4 4QL, United Kingdom.}
\affiliation{Living Systems Institute, University of Exeter, Stocker Road, Exeter, Devon, EX4 4QD, United Kingdom.}

\date{\today}

\begin{abstract}
The chiral(ity)-induced spin selectivity (CISS) effect, where electrons passing through a chiral medium acquire significant spin-polarization at ambient temperatures, has been widely observed experimentally, yet its theoretical foundations remain actively debated. Open questions persist regarding whether CISS originates from helical geometry or more general chirality, and whether a unified mechanism can account for phenomena across solid-state and soft-matter systems, mesoscopic films, and single molecules. Clarifying the interrelations between existing models is essential to determine if a universal picture of CISS can be found or whether system-specific models are required, and if so, where their common starting point should lie for a workable classification of CISS manifestations. Despite this theoretical fragmentation, recent studies of CISS effects in electron transfer systems, magnetic field sensitivity and coherence of radical pair reactions, polarized electroluminescence in chiral hybrid perovskites, DNA-based biosensors, and enantioselective detection, highlight its broad conceptual relevance and potential applications in spintronics, molecular sensors, and quantum information processing. In this review, we help bridge the gap between theory, experiment, and implementation, with a particular focus on prospects for quantum sensing and metrology. We outline fundamental frameworks of CISS, clarifying what constitutes the `chiral', the `induced', and the `spin-selectivity' that makes up CISS, before going on to survey key model realizations and their assumptions. We examine some of the emerging quantum sensing applications and assess the model-specific implications, in particular exemplifying these in the context of spin-correlated radical pairs, which offer a promising, tunable, and biomimetic platform for emerging molecular quantum technologies.  
\end{abstract}

\maketitle

\section{Introduction: On Chirality and Chiral(ity)-Induced Spin Selectivity}
\label{part1}

The notion of chirality stems from fundamental concepts in geometry. Roughly, an object is chiral if it breaks mirror symmetries along any plane of reflection. In high-energy physics, the Wu experiment famously demonstrated $\beta$-decay to be spin-selective and chiral (in Poincar\'{e} transformations), establishing how the weak interaction violates parity \cite{wu57}. Chirality has since come to play a pivotal role across scientific disciplines, for instance in biochemistry---where enantiomers (i.e. chiral images) of the same compounds, that otherwise ought to be chemically identical, can exhibit significant pharmacological differences due to differing chirality \cite{vargesson_thalidomide-induced_2015, nguyen_chiral_2006}. By definition, chiral biomolecules can exist in all their enantiomeric configurations; yet in-vivo, all genetic polymers (DNA/RNA, and their composite residues) exclusively exist in their right-handed enantiomeric forms, while all amino acids are found in the left-handed form. This gives rise to the emergence of self-assembling chiral structures on a macromolecular and cellular scale \cite{liu_supramolecular_2015}. Thus, chirality arguably arises as a defining feature of life itself, with living systems by and large exhibiting almost perfect homochirality despite the significant entropic costs entailed \cite{blackmond_origin_2019, chen_origin_2020, ozturk_origin_2023} (for interesting exceptions, see a review on chirality and extraterrestrial life by Glavin \textit{et al.} \cite{glavin_search_2020}). While such asymmetry strongly suggests that chirality plays a pivotal role in regulating biochemical processes, the precise mechanisms by which it does so remains a long-standing scientific mystery \cite{ozturk_origin_2023, ozturk_central_2023}. A prominent candidate to address this gap in our understanding is the phenomenon of chiral(ity)-induced spin-selectivity (CISS), which broadly encompasses how electrons passing through a chiral (and often helical) medium acquire spin-polarization, and other anomalous spin dynamics \cite{Megan2025_analogSimul, bradbury2025, aharony2025, glass25, Chen2024, activeglassy24, fransson_charge_2022, shitade_geometric_2020}. 

The CISS effect was first observed, and the term introduced, by Ray \textit{et al.} in 1999 \cite{ray_asymmetric_1999}, when spin-polarized photoelectrons ejected from thin chiral biomolecular films adsorbed onto an Au interface displayed asymmetric scattering $\sim10^3$ larger than those reported in chiral gas-phase studies \cite{Mayer_dichroism_1995, Nolting_dichroism_1997}, suggesting for the first time a connection between helical geometry and quantized spin angular momentum. Subsequent studies on biomolecules have observed CISS to be optimized at room temperatures \cite{sang_temperature_2021, das_temperature-dependent_2022}, thus raising exciting possibilities for designing robust chiral(ity)-based quantum devices, and spintronic applications, operating at ambient conditions. Experimental studies of CISS, seeking to uncover its fundamental underpinning mechanisms, have since flourished into a dynamic and multifaceted field. It has furthermore emerged as a spin-manipulating paradigm underpinning an opportunity for designing controllable and precise quantum sensors \cite{lin_SCRP_Qubits_2025, ciss24spin, zakrzewski24, mani_review_2022, aiello2022, vizvary2021quantum, cohen2011}. Overall, the key experimental realisations of CISS can be broadly partitioned into four broad categories:
\begin{enumerate}
    \item \textbf{Spin-to-charge conversion through metal-to-chiral-medium junctions}, including microscopic studies of driven electron transport through chiral biomolecular nanojunctions \cite{Aragones2017, Singh2025, Ortuno2023}, quantum Hall-response \& induced magnetic order of metal interfaces with adsorption of chiral monolayer \cite{kumar_chirality-induced_2017, ben_dor_magnetization_2017, meirzada_long-time-scale_2021, liu_magnetless_2024, moharana_hall_2025}.

    \item \textbf{Photoexcitation of chiral molecules}, in particular photoemission spectroscopy and Mott polarimetry of chiral monolayer absorbed onto metal substrate \cite{ray_asymmetric_1999, Rosenberg2015, abendroth_spin-dependent_2019}.

    \item \textbf{Interface-less biochemical processes}, including intersystem crossing dynamics \& radical pair formation in donor-chiral bridge-acceptor enantiomers \cite{eckvahl_direct_2023}, and arguably chiral transmission \cite{mondal_spin_2021, pandey_chirality_2022}. 

    \item \textbf{Application-driven and biomimetic device development}, which seeks to harness the robustness of CISS at ambient temperatures. These include emerging technologies for efficient enantiomeric separation \cite{banerjeeGhosh2018, banerjeeGhosh2020, Metzger2020}, oxygen-evolution \cite{Mtangi2017, zhang2018}, molecular machines \cite{kim2021}, molecular qubits\cite{lin_SCRP_Qubits_2025}, spin-optoelectronics \cite{He2025, pan_spinOpto, yang_spinOpto, zhang_spinOpto, tang_spinOpto}, quantum-assisted magnetic sensing \cite{volker2023, giaconi2024, BANGRUWA2023133447}, and prospectively room-temperature qudits \cite{aiello2022}.   
\end{enumerate}

\begin{figure}[h]
\includegraphics[width=7cm]{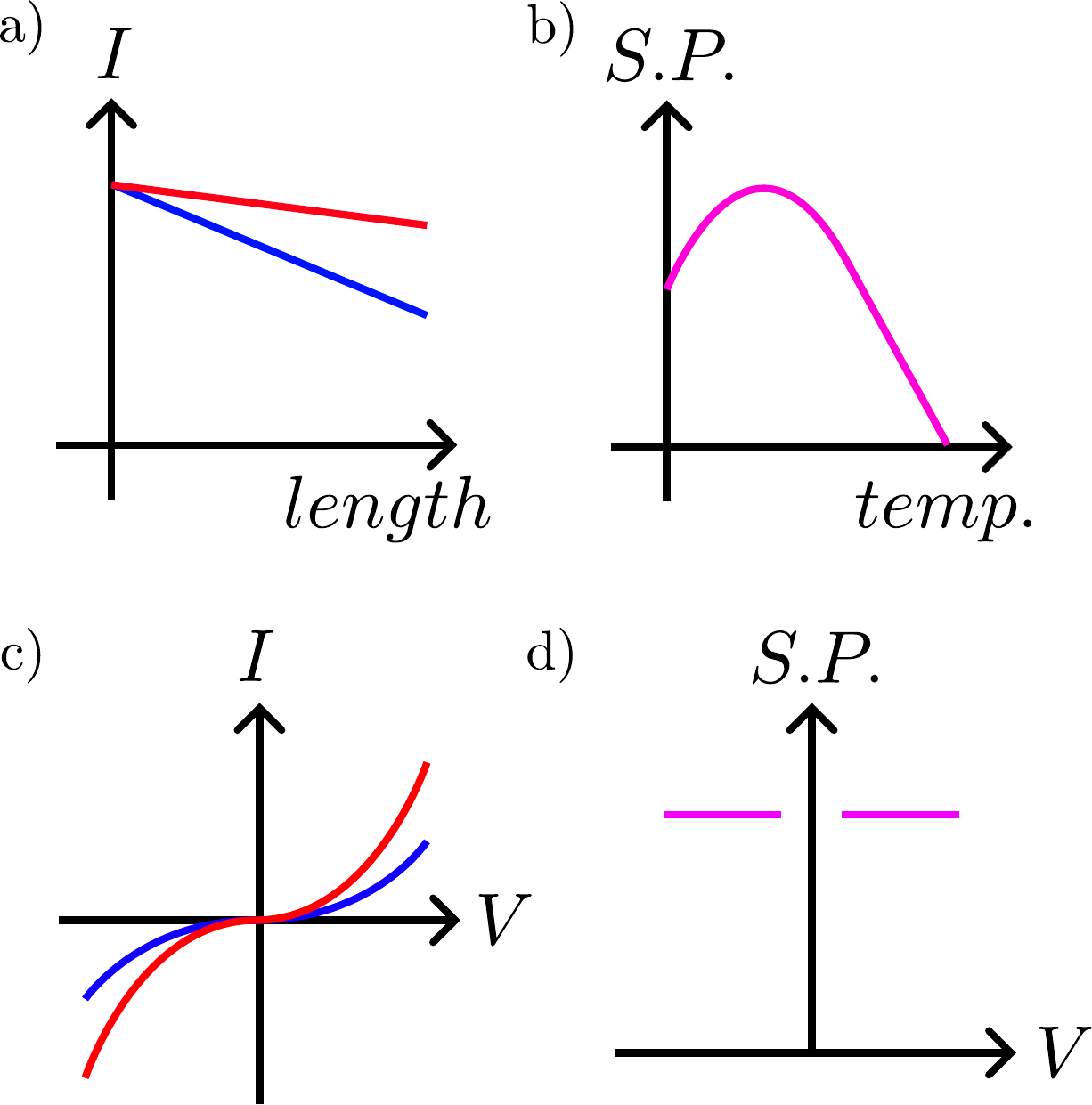}
\caption{Example graphs for structure-property relations used to characterize CISS at steady-state for spin-valve-like setups in which an organic chiral molecule or self-assembled monolayer acts as the spacer: the blue and red lines represent opposite magnetization of the spin-analyzer (for example, an atomic-force-microscopy tip); (a) charge current may decrease at a higher rate with respect to length of the chiral layer when setup magnetization is unfavorable to majority spins under transport ; (b) spin-polarization as indexed by magnetocurrent anisotropy may peak at near-ambient temperatures; (c) charge current-potential difference graphs may suggest nonlinear-response and point to charging effects or the dielectric behavior of the chiral organic layer being strong; (d) spin-polarization may be relatively constant with regards to bias voltage, save for an undervolted regime (in which too little current passes for reliable measurement) and an overvolted regime.}
\label{fig:figure1}
\end{figure}

\begin{figure}[h]
\includegraphics[width=\columnwidth]{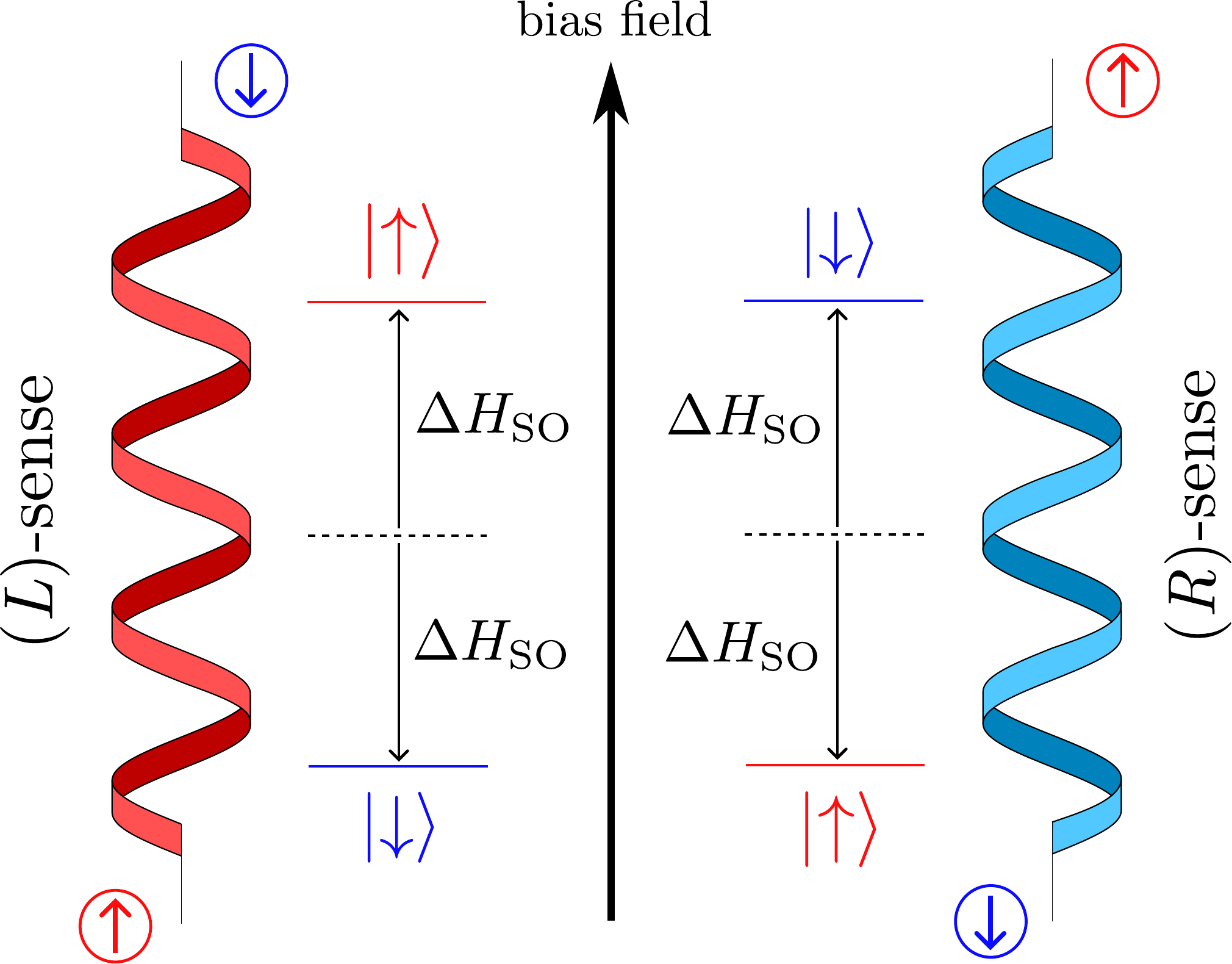}
\caption{Schematic diagram of the naive CISS mechanism. Electrons moving
along a helical potential will experience an effective magnetic field that
interacts with the electrons’ spin in a manner such that opposite spin-orientations will prefer to travel in opposite directions. This spin-preference flips when the helical-configuration is reversed from left-handed to
right-handed and vice-versa.}
\label{fig:figure2}
\end{figure}

While the above does not serve as an exhaustive overview, it lays out broadly illustrative examples on the state of the field, which has covered wide ground since its inception. This diversity of experiments, though telling of the universality of CISS, has also contributed to a highly-fragmented picture of CISS characteristics. What often emerges is a set of structure-property/function relationships (Fig.~\ref{fig:figure1}), that are not directly comparable across experiment classes \cite{pyurbeeva23}. This intriguing breadth of findings calls for theoretical efforts to better account for and predict CISS properties. To this end, a spin-momentum locking picture is often employed, proposing spin-orbit coupling (SOC) to be the defining component in CISS. In this framework, SOC couples the linear momentum of an electron to its spin angular momentum, resulting in electron motion being energetically favorable when they are parallel or antiparallel, depending on the handedness of the helical medium (Fig.~\ref{fig:figure2}).

A SOC mediated mechanism alone, however, cannot account for the spin-selectivity magnitudes observed in experiments \cite{naaman_chiral_2020, naaman2012, naaman2015, Michaeli_2017}, which becomes apparent from an elementary order-of-magnitude \cite{aiello2022} estimation. Consider the rest-frame of an electron in motion along a helical path with some incident electric field, $\mathbf{E}_{\mathrm{ext}}$: the electron incurs an effective magnetic field due to the field, $\mathbf{B}_{\mathrm{eff}} = -c^{-2} \vec{v} \times \mathbf{E}_{\mathrm{ext}}$, while possessing a magnetic moment, $\vec{\mu} = -g_{S}\mu_{B}\hbar^{-1} \mathbf{S}$. The effective magnetic field exerts a torque on the magnetic moment, resulting in a Larmor interaction energy that is equivalent to a Rashba-like spin-orbit coupling: $\Delta H_{\mathrm{SO}} = -\vec{\mu} \cdot \mathbf{B}_{\mathrm{eff}} = \lambda \hat{n} \cdot \left( \vec{\sigma} \times \vec{p} \right)$. By this account, spin-degeneracy is broken and an energy gap opens. Assuming Boltzmann statistics, the energy gap for this spin-splitting gives rise to a population ratio to the factor of $\exp \left( -2\Delta H_{\mathrm{SO}}/{k_B T} \right)$. Supposing one expects to retrieve a 20\% polarization at ambient temperatures ($\sim$300\,K), this requires an expected spin-splitting field strength of $\sim$5\,meV, which is magnitudes apart from expected SOC splitting energies of bare hydrocarbons. Much activity in CISS theory has since been devoted to accounting for the anomalous spin-polarization strengths observed. Concurrently, models attributing CISS’s strength to rich nonequilibrium interfacial effects have also been proposed, from interface-orbital magnetization \cite{alwan_spinterface_2021, dubi_spinterface_2022} to spin-blockade \& charge-displacement mechanisms \cite{naaman_chiral_2020, fransson_charge_2022}, collectively giving rise to a more comprehensive theoretical treatment of CISS in nanojunctions. 

It is increasingly evident that CISS should not be treated as a single, homogeneous phenomenon, but rather as a set of related effects that manifest differently depending on the experimental context. Each regime---be it solid-state transport, soft-matter photophysics, or biochemical electron transfer---possesses its own assumptions and constraints, requiring tailored theoretical approaches. However, this fragmentation has in turn given rise to a vast array of frameworks that preclude inter-applicability. Given the diversity of materials and platforms in which CISS finds itself relevant, a fundamental question is raised: What lies at the core of CISS? It is our view that, contrary to the perceptible fragmentation in the theoretical literature, a coherent and unifying understanding can nevertheless be gleaned. In this review we proceed by first laying out brief summaries of core mechanisms proposed to underlie CISS. Next, we consolidate developments in the theoretical literature in a systematic manner to give a brief overview of the trends thus far and highlight future directions. We conclude with a discussion of CISS's technological potential in the context of quantum sensing. We note that this review is not intended as a comprehensive survey of the CISS literature---broader accounts can be found elsewhere \cite{bloom_chiral_2024}---but instead we focus on discerning common identifiable conceptual threads across the diverse manifestations of CISS, the emergence of distinct mechanisms in specific experimental realizations, and the implications these have for emerging quantum metrology applications \cite{boeyens2025, lee2025tunable, lin_SCRP_Qubits_2025}. We hope this contribution will assist in demystifying CISS and underscore its potential to be primed for near-term applications in spintronics, biochemical industrial application and the quantum information sciences \cite{chir_gates25, chiesa2025ciss, kermiche22}. 

\section{How Does Spin Selection in CISS Occur?}

Although CISS is conceptually underpinned by spin-transfer, to our knowledge there currently exists no apparatus that can directly measure spin-currents. Observation of CISS thus necessarily relies on spin-to-charge-current conversion and must draw from more indirect signatures. Benchmarking of CISS is further compounded in difficulty by the soft-matter and dielectric properties of biomolecular thin films giving rise to potentially inconsistent structure-function properties of unclear origin, most exemplified by the temperature-dependence of CISS \cite{carmeli_spin_2014, das_temperature-dependent_2022, sang_temperature_2021, alwan_temperature-dependence_2023}. A more thoroughgoing review of the experimental landscape and the many ways CISS has been indexed can be found in Bloom \textit{et al.}'s extensive review \cite{bloom_chiral_2024}. Theoretical investigations in CISS-junction transport have in turn been split between using either spin-transmission probabilities or charge-current values as a measure of spin-polarization, which are not equivalent nor necessarily related quantities \cite{fransson_chiral_2022}. Additionally, the notion that spin-selectivity alone is being measured is called into question by Liu \textit{et al.} \cite{liu_chirality-driven_2021}, who notes that spin no longer provides a good quantum number due to SOC; CISS rather selects via the total electronic angular momentum quantum number ($j = s + l$), suggesting potential for orbitronics applications. This mix of analyses brings to the fore the proverbial elephant-in-the-room: how does spin-selection in CISS precisely occur, and how should experimental signals be understood in the context of proposed mechanisms? In the subsections to follow, we sequentially cover the questions: What is spin-selectivity? How is it induced? And what falls within the purview of `chirality' in CISS?

\subsection{Does Spin-Orbit Coupling Alone Give Rise To Spin-Selectivity?}
\label{part2A}

Insofar as SOC is pivotal to CISS, the question of how SOC affects spin-selection is seldom addressed as compared to how SOC may have been enhanced in CISS to close the strength-gap. The spin-momentum locking picture posits simply that electrons with unfavorable spin-states relative to their motion will back-scatter, but by what means will this back-scattering occur? Indeed, Bardarson's theorem, which follows on from Kramers' degeneracy, implies that for time-reversal symmetric non-dissipative spin-1/2 systems, the transmission eigenvalues of both spin-configurations should be equal (i.e. spin-polarization should be forbidden in two-terminal junctions) \cite{Bardarson_2008}; its implications for CISS have been a chief subject of exploration by Utsumi, Aharony, \& Entin-Wohlman \cite{aharony2025, utsumi_2022,utsumi_spin_2020}. 

How, then, can we reconcile the above mentioned fundamental symmetry constraint with the experimentally observed spin selectivity in chiral systems? Let us first examine the equilibrium case in which no external bias is applied onto our system. We start by detailing a more commonly-encountered perspective from spintronics: spin-orbit coupling gives rise to spin-momentum locking, which can arise in spin-polarized currents. Considering a single-channel model, a Rashba-like spinor Hamiltonian for a quantum wire can phenomenologically describe the essential physics of our system, which can be denoted in reciprocal-space by:
\begin{equation}
    \hat{H}(k) = -2t \sigma_0 \cos{k} + 2\alpha \sigma_z \sin{k} - h \sigma_x
\end{equation}
where $t,\alpha, h \in \mathbb{R}$, while $\{\sigma\}$ denote standard Pauli matrices.

\begin{figure}[h]
\includegraphics[width=8cm]{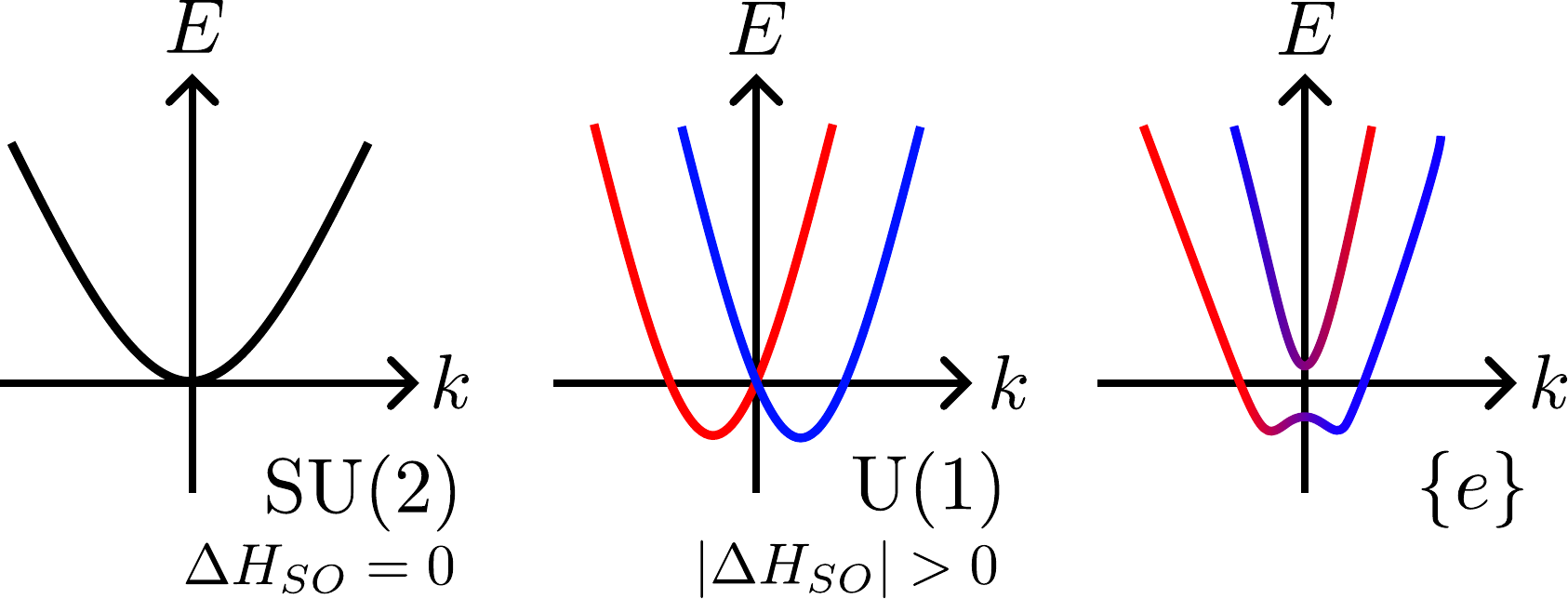}
\caption{Dispersion relations for 1D Rashba-like Hamiltonian. The spin energy-bands are degenerate before the introduction of SOC (left). At weak SOC strengths, the bands split (middle) while spin-spin correlations give rise to an avoided crossing (right), describing how spin-momentum locking occurs. Notably, one expects the electronic ground-state to have non-zero crystal-momentum and non-zero spin angular-momentum.}
\label{fig:fig3}
\end{figure}

Visualising the spin-state energy bands in Fig.~\ref{fig:fig3}, finite $\alpha$ describes SOC effects and results in the splitting of the spin energy-bands (note that these are not spin-polarized); and a further non-zero $h$ denotes spin-spin correlations or quasi-Zeeman effects that mix the spin-bands and gives rise to avoided crossings. Here, it is apparent that SOC lifts spin-degeneracy (specifically breaking SU(2) symmetry) and gives rise to spin-momentum locking. Intuitively, one would expect that the helical geometry \textit{intrinsically} gives rise to a finite spin-current. And yet, it would appear that such an expected spin-current is not sufficient in arriving at an asymmetric spin-accumulation.
\begin{figure}[h]
\includegraphics[width=7cm]{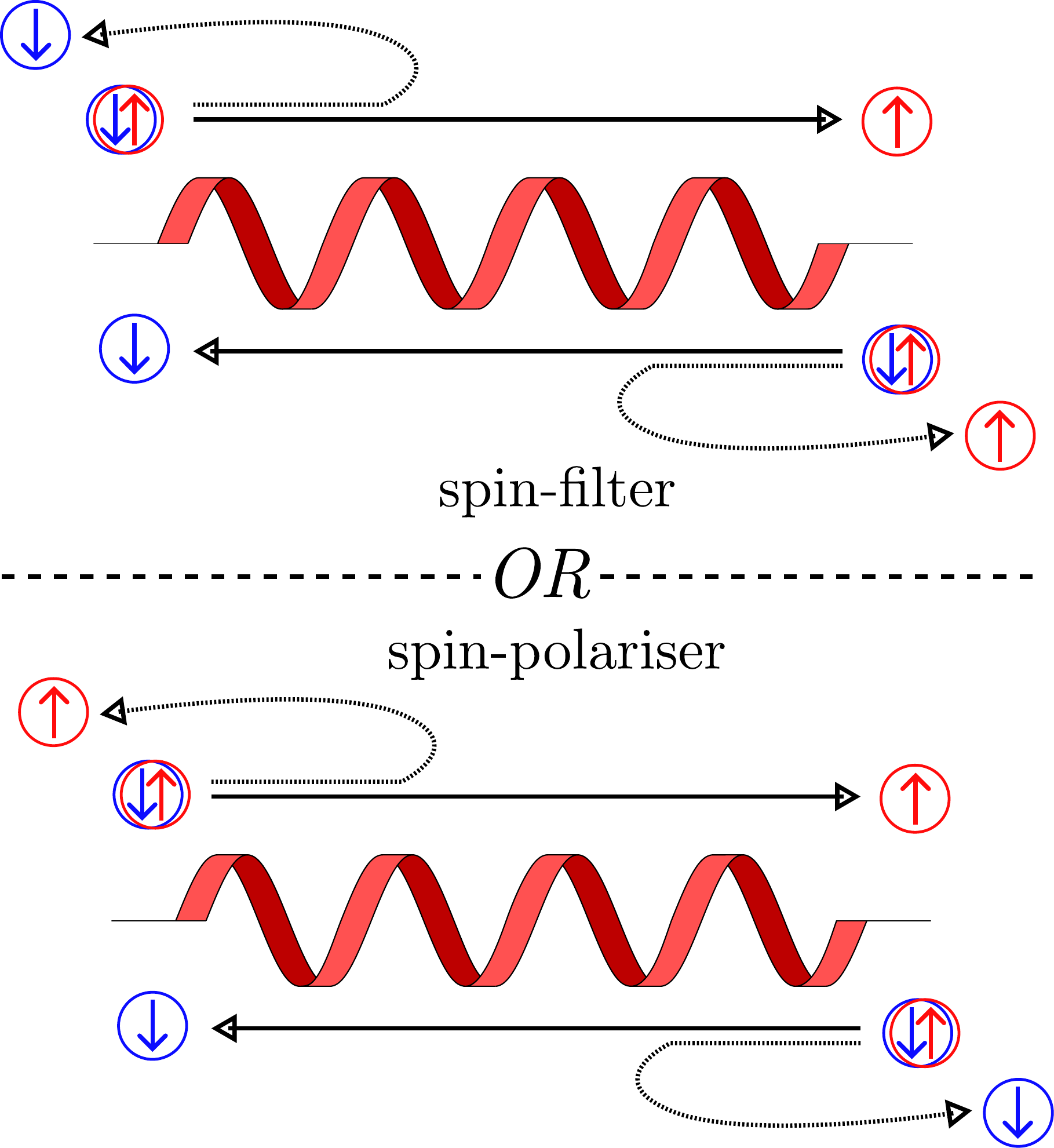}
\caption{Schematic diagrams of spin-filter (top) and spin-polarizer (bottom) pictures. In the spin-filter picture, spin-momentum locking results in spin-dependent transport such that opposite-valued spins are preferentially transmitted in opposite directions. Spinful particles travelling in unfavorable directions are reflected. In the spin-polarizer picture, reflected particles necessarily undergo a spin-flip since no dissipative spin transport can occur at equilibrium.}
\label{fig:fig4}
\end{figure}
We can consider an ideal scenario of CISS exhibiting dissipationless coherent/ballistic transport with unpolarized sources. In the absence of an external applied magnetic field, time-reversal symmetry (of the Hamiltonian) is preserved; Wolf \textit{et al.} \cite{wolf_unusual_2022} points out that no equilibrium spin transport or accumulation can arise accordingly, since equilibrium dictates the scattering-matrix relations: 
\begin{equation}
    \sigma_{r} + \sigma_{t'} = 0 \quad ; \quad \sigma_{r'}+\sigma_{t} = 0,
\end{equation}
in which $\sigma_{r(t)}$ denotes the spin expectation-value for the reflected(transmitted) wave, while primed variables correspond to those of an incident wave entering from the opposite end. 

Given $\sigma_{t} = -\sigma_{t'}$, for injected spin currents, back-scattered electrons must have the same spin as transmitted ones:
\begin{equation}
    \sigma_{r} = \sigma_{t} = -\sigma_{r'} = -\sigma_{t'}.
\end{equation}
In other words, reflected waves and transmitted waves incident from opposite directions, hence travelling in the same direction, hold opposite spin-states (Fig.~\ref{fig:fig4}), a conclusion previously corroborated by a scattering-matrix analysis from Utsumi \textit{et al.} \cite{utsumi_spin_2020}. Additionally, we may argue for this on \textit{ad absurdum} grounds: under the spin-momentum locking picture, a lone helical molecule at finite temperature would undergo spin-accumulation on its ends by itself owing to Brownian electron motion at equilibrium and become fully self-polarized barring any relaxation mechanisms; so all coiled wires would be spin-polarized by default. But this alone does not suggest a physical basis as to how such spin-selection may arise. Yang \textit{et al.} \cite{yang_spin-dependent_2019, yang_reply_2020} proposes along similar symmetry grounds that a spin-flip reflection must be inherent to CISS; but assuming ballistic or dissipationless travel, no scattering events in general can occur within the CISS molecule to give rise to such a spin-flip. One possible resolution to this is interfacial: the CISS-junction itself acts as a scatterer that induces spin-flip scattering prior to entry. Previous spin-active tunnelling studies \cite{varela_spin-orbit_2020, michaeli_origin_2019} have proposed that the helical shape of the junction generates a curvature-induced quantum geometric potential that causes spin-flips. 

\begin{figure}[h]
\includegraphics[width=7cm]{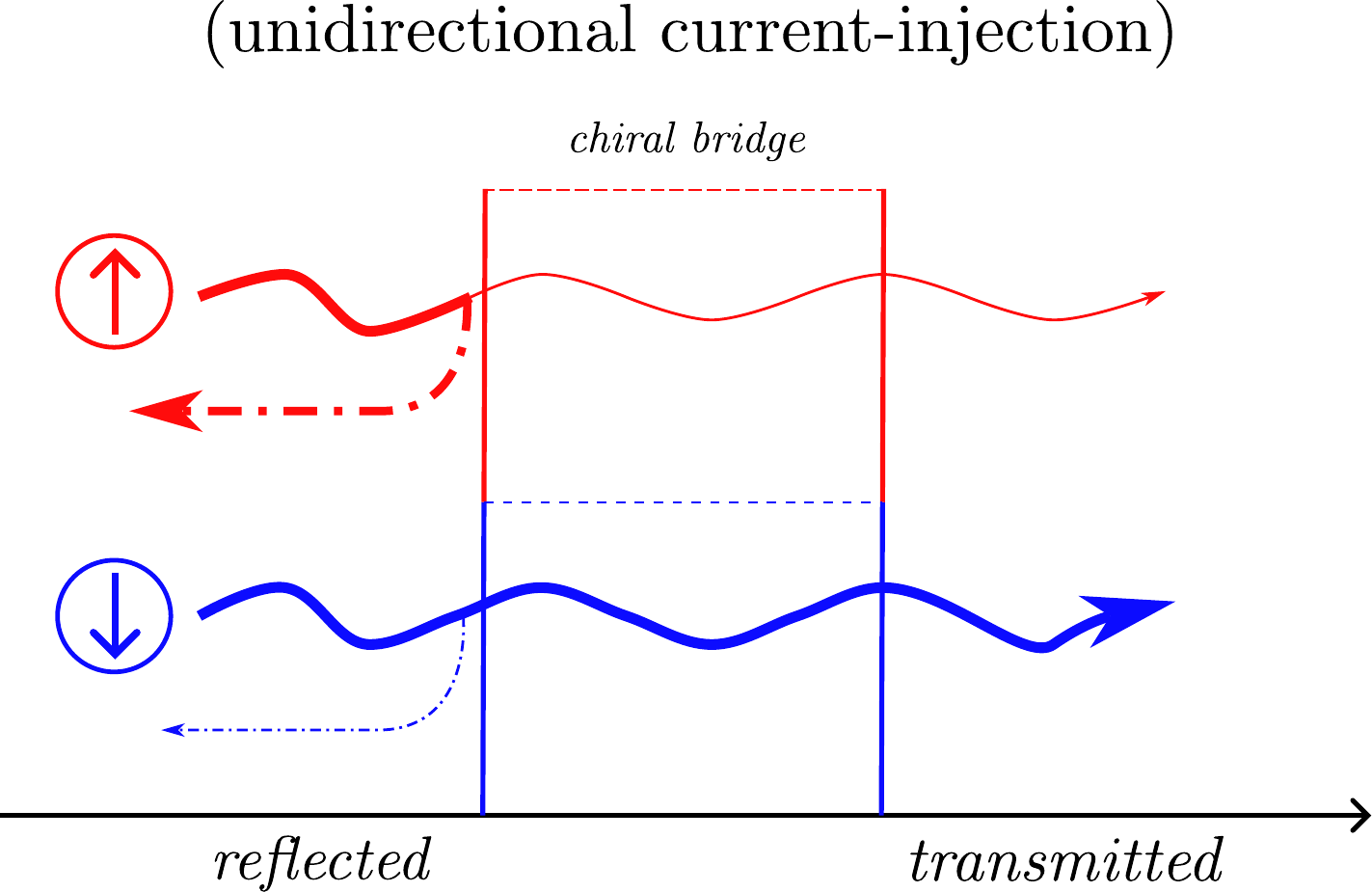}
\caption{Illustration of CISS as a barrier potential, showing unidirectional charge current. As an example, spin-up particles have to tunnel through a larger barrier and are thus preferentially reflected. In the equilibrium case, when bidirectional current is considered, this will result in zero spin transport and accumulation. Thus, the `inherent spin-flip' necessary for the spin-polarizer picture can be construed as an interfacial effect.}
\label{fig:fig5}
\end{figure}

Alternatively, minimal models recast the CISS-junction as a scattering region with an anisotropic dipole field \cite{ghazaryan_analytic_2020} or treat the SOC as a barrier potential varying with direction as studied by Varela \textit{et al.} \cite{varela_spin-orbit_2020} as illustrated in Fig.~\ref{fig:fig5}, though it has been noted that the latter effort does not confer spin-selectivity as-is \cite{wohlman_comment_2021}. Additionally, we may extend from the equilibrium regime into the transport regime and view CISS as a form of current-induced spin-polarization. 
Suppose one subjects a linear electric bias onto a Fermi distribution of electrons within such a picture (i.e. $\exists E \Rightarrow f_k \rightarrow f_k^{0} + \delta f_k$). This gives rise to a non-equilibrium steady state in which the spin states are split, giving rise to a spin-asymmetry describable as:
\begin{equation}
    \delta s = \sum_k \braket{S}_k \delta f_k,
\end{equation}
where $\braket{S}_k$ denotes the expectation value of spin for the $k$-th mode. Invoking the Boltzmann approximation, we can recast the spin current as a spin-wave:
\begin{equation}
    \delta s = \sum_k \braket{S}_k \gamma^{-1}_k (v_k \cdot E)  \frac{\partial f^0_k}{\partial \varepsilon_k},
\end{equation}
where $\gamma_k$ indicates the scattering rate, $\varepsilon_k$ the band energy, and $v_k$ the spin-wave group velocity of the \textit{k}-th mode. The electric bias also generates a charge current given by:
\begin{equation}
    j = -\frac{e}{V} \sum_k v_k \delta f_k,   
\end{equation}
and as such, so long as the spin expectation-value is non-zero for some $k$, it would seem that a charge current is naturally accompanied by a finite spin current. However, is such spin current tantamount to spin-transport or spin-accumulation? Even in the transport regime, it can be shown that such a shifting of the Fermi sea is not equivalent to \textit{preferential transmission of spin-species}. We can call upon a physical argument by taking a gauge-covariant formulation of the Hamiltonian with SOC and obtaining its effective electromagnetic fields, an approach initially offered by Fr\"{o}hlich \& Studer \cite{FrohlichStuder1993} and Tokatly \cite{Tokatly2008}. Consider first the generic Pauli-Schr\"{o}dinger Hamiltonian for a lone electron in an external scalar potential $\pmb{\Phi}$ and vector potential $\mathcal{A}$, with corrections up to $c^2$ (derivable from performing the Foldy-Wouthuysen transformation onto the Dirac Hamiltonian):
\begin{equation}
    \begin{split}
        H &= \frac{\pmb{\Pi}^2}{2m_0} + m_0c^2 -e\pmb{\Phi} + \frac{e\hbar}{2m_0 c} B \cdot \sigma + \frac{e\hbar^2}{8m_0^2c^2} \mathrm{div}\> E \\
        &\quad + \frac{e\hbar}{8m_0^2c^2} [\pmb{\Pi} \cdot (\sigma \times E) + (\sigma \times E) \cdot \pmb{\Pi} ],
    \end{split}
\end{equation}
in which $E=-\nabla\pmb{\Phi} - (1/c) \partial_{t}\mathcal{A}$ is the electric field experienced, $B = \mathrm{curl}\> \mathcal{A}$ is the magnetic field experienced, $\pmb{\Pi}$ is the canonical momentum, defined by $\pmb{\Pi} = -i\hbar\nabla + (e/c)\mathcal{A}$. In order of appearance, the Hamiltonian includes contributions from the electron's kinetic energy, its rest mass, potential energy due to $\pmb{\Pi}$, the Zeeman term, the Darwin term, and lastly the spin-orbit coupling term.

The Pauli equation hosts a U(1) x SU(2) symmetry group; since the electromagnetic U(1) gauge field and the isospin SU(2) gauge fields commute, we may safely focus only on how SU(2) transforms. Using Einstein summation convention, let us define $\mathcal{A}_\mu$ ($\mu \in {0,x,y,z}$) as the non-Abelian SU(2) vector potential such that its components are given by:
\begin{align}
    \mathcal{A}_{0} &= -\frac{e\hbar}{m_0 c} B^{\nu} \tau^{\nu}; \quad \nu\in[x,y,z],\\  \mathcal{A}_{i} &= -\frac{e\hbar}{m_0 c^2} \varepsilon_{ij\nu} E^{j} \tau^{\nu}; \quad i,j\in[x,y,z],
\end{align}
where $\varepsilon$ is the Levi-Civita symbol and $\tau^{\nu} = \sigma^{\nu}/2$ are the generators of SU(2) in the fundamental representation (i.e. normalised Pauli matrices). By Noether's theorem, the gauge invariance of our Hamiltonian implies the existence of a covariantly conserved current, $\xi$, governed by: 
\begin{equation}
    D_\mu \xi_\mu = 0,
\end{equation}
where $D_\mu$ represents the covariant derivative set as $D_\mu = \partial_\mu -i \mathcal{A}_\mu$. The definition of covariant derivatives allows us to form a compact Euler-Lagrange equation after recasting the Hamiltonian operator as $i\partial_{t}$:
\begin{equation}
    i\hbar D_0\psi=-\frac{\hbar^2}{2m_0}\sum_{i} D_{i} D_{i}\psi,
\end{equation}
where $\psi$ is the fermionic field operator, $(\psi_{\uparrow}, \psi_{\downarrow})$. This lends to a simple formulation of the action:
\begin{equation}
    S=\int \mathrm{d}t \, \mathrm{d}\mathbf{r} \>\> i\hbar\psi^{\dagger}(D_0 \psi) - \frac{\hbar^2}{2m_0} \sum_{i}(D_i \psi)^{\dagger}(D_i \psi),
\end{equation}
where we have implicitly evaluated the wavefunction in position space. Here, it can be shown that the action remains gauge-invariant under local transformation, definable as $\Psi \rightarrow U\Psi, \>\> \mathcal{A}_\mu \rightarrow U\mathcal{A}_\mu U^{-1} - i(\partial_{\mu}U)U^{-1}$, where $U=e^{i\theta^\alpha \tau^{\alpha}}$ is some arbitrary unitary SU(2) matrix. And taking the matrix decomposition, $\xi_{i} = \xi^{\alpha}_{i}\tau^{\alpha}$, such that $\xi^{\alpha}_{i} = \partial{S}/\partial{A^{\alpha}_i}$, the form of the Noether current can thus be written out as:
\begin{align}
    \xi^{\alpha}_{0} &= \psi^{\dagger}\tau^{\alpha}\psi = s^{\alpha}, \\
    \xi^{\alpha}_{i} &= -\frac{i}{2m_0}\left[ \psi^{\dagger}\tau^{\alpha}(D_i \psi) - (D_i \psi)^{\dagger}\tau^{\alpha}\psi\right] - \frac{1}{4m_0} A^{\alpha}_i \hat{n},  
\end{align}
where $\hat{n}$ is the unit-vector in the $i$-direction. We see explicitly that $\xi_0$ corresponds to the spin density; accordingly, $\xi_i$ returns the spatial spin-current we are after.

The above also leaves us with the continuity equation:
\begin{equation}
    \partial_t \xi^{\alpha}_\mu + \varepsilon_{\alpha \beta \eta} A^{\beta}_\mu \xi^{\eta}_{\mu} = 0,
\end{equation}
in which the second term encodes the chiral anomaly via spin precession in a magnetic field for $\mu = 0$, and otherwise an ``internal spin torque"\cite{Tokatly2008} for $\mu =x,y,z$. As the current form of the spin current can be cumbersome, one can instead derive the SU(2) effective field tensor:
\begin{equation}
    F_{ij} = \partial_{i} A_{j} -\partial_{j}A_i -i[A_i,A_j],
\end{equation}
Here, $j=0$ gives the dissipative electric-like part of the field while $j=x,y,z$ gives the nondissipative magnetic-like part of the field. The effective magnetic field gives rise to non-dissipative equilibrium spin currents and hence do not contribute to spin accumulation \cite{GebauerCar2004}; contributions to spin accumulation must arise from the dissipative effective electric field.

If no external magnetic field is applied, then $F_{i0}$ is reduced simply to $-\partial_t A_i$. Supposing a simplified CISS model in which transport is set in the $z$-direction, we may ignore $F_{x0}$, $F_{y0}$ and we assume that only $\partial_tE_z$ is negligible, giving:
\begin{equation}
    F_{z0} = \frac{e\hbar}{m_0 c^2} \partial_t(E^x \tau^y - E^y \tau^x).
\end{equation}
In other words, spin-up and spin-down states experience the same dissipative field strength up to a phase difference. On the contrary, the non-Abelian effective magnetic fields (such as $F_{xy} = \frac{e\hbar}{m_0 c^2} ( |E^x| - |E^y|) \tau^z $) may display unequal field strengths between spinor states. This generates a spin current that is non-dissipative and hence does not contribute in scattering-matrix analyses. Unlike charge currents, non-dissipative spin current can still persist due to chiral anomaly even if no dissipative spin current can be drawn, and this can result in a magnetic order or spin-texture when coupled with spin-momentum locking, such that biomolecular CISS under bias effectively gives rise to soft matter ferromagnets. Such magnetisation is linked to spontaneous time-reversal symmetry breaking, and is not to be identified with explicit breaking of time-reversal symmetry at the level of the Hamiltonian, which remains invariant \cite{zollner_insight_2020}. It is thus difficult to take spin-selectivity as an intrinsic consequence of helical geometry. If the spin-filter picture is to be applicable, CISS should arise from scattering and external dissipative factors \cite{guo_spin-dependent_2014, fransson_chiralOrigin_2025}, interfacial effects, and charge-trapping \cite{zhao25} etc. as will be discussed in Sec. \ref{part3E}. 

Non-dissipative and/or pure spin-currents may be simply probed by spin-to-charge conversion using the inverse spin-Hall effect, the Hanle effect, or by way of time-resolved Kerr microscopy, as demonstrated by Sun \textit{et al.} who showcase CISS giving rise to highly anisotropic nonlocal Gilbert damping \cite{sun_colossal_2024}.  However, as experimental setups often stage chiral molecules/molecular layers either in contact with magnetised materials or within external magnetic fields, CISS is commonly observed in inherently non-equilibrium conditions which the idealised scenarios described in this sub-section may be far removed from. Such insight is nevertheless essential to understanding the phenomenon and the directions future applications may take; we leave discussion of this rich topic for last in Sec. \ref{part4}.

\begin{figure}[b]
\includegraphics[width=\columnwidth]{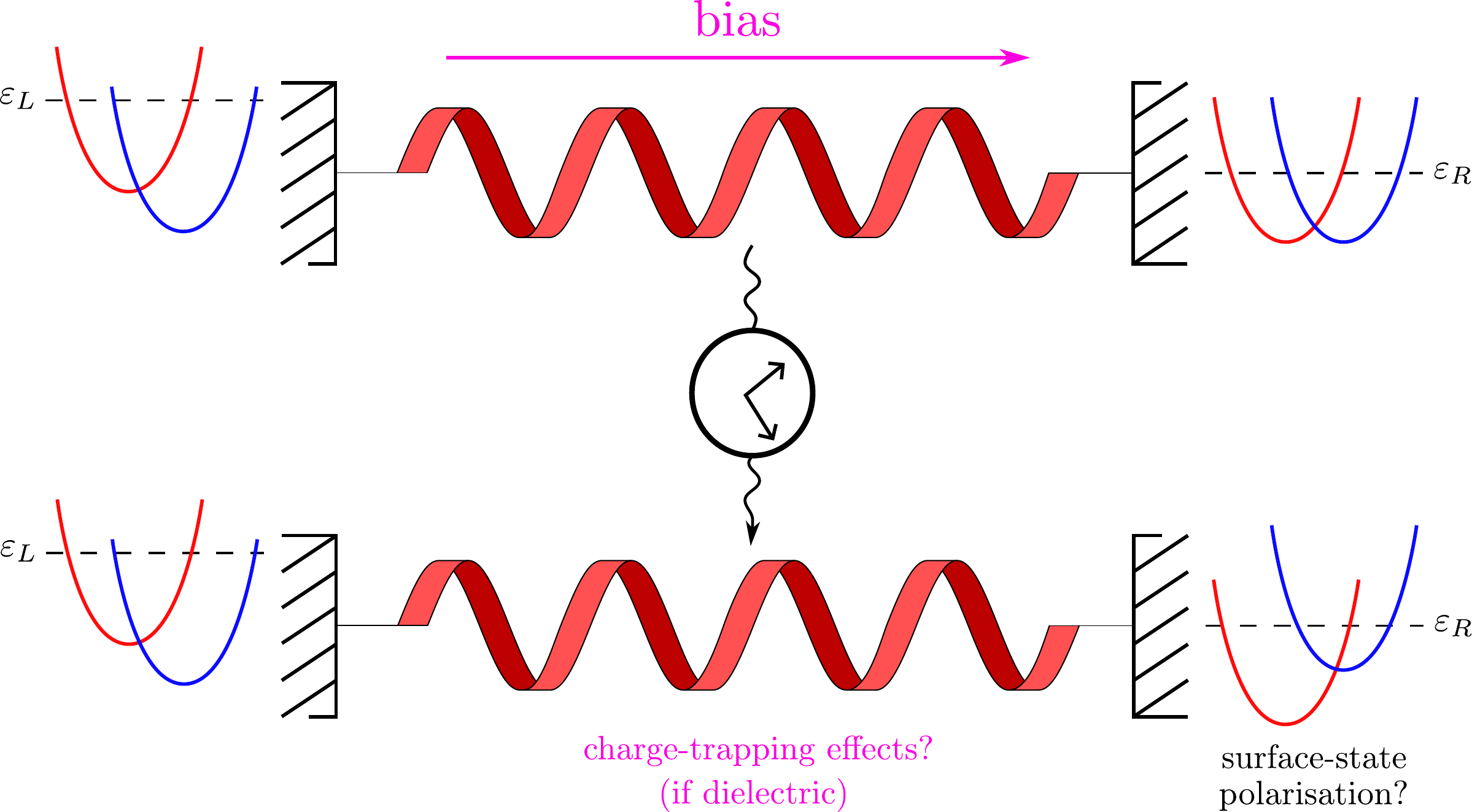}
\caption{Electric bias application across the biomolecular nanojunction can change the electronic properties of the setup-at-large, giving rise to surface-state polarization, charge-trapping, and other effects contributing to nonlinear-response.}
\label{fig:fig6}
\end{figure}

\subsection{Onsager's Relations: Linear or Non-linear? Molecular vs Interfacial Mechanisms?}
\label{part2C}

A persistent question in the CISS corpus is whether anomalous spin-selectivity strengths as reported in experiments originate intrinsically from SOC or whether these are enhanced by medium-specific processes. In other words, can CISS exist as a linear-response phenomenon? Briefly, given an out-of-equilibrium system in the linear regime, for which the coupled flux-force transport equations are \(J_j = \sum_k L_{jk} \mathcal{F}_{k}\) with fluxes \(J_j\) and generalised forces \(F_k\), Onsager's reciprocal relations state that the kinetic coefficients \(L_{jk}\) obey \(L_{jk}=L_{kj}\) (i.e. they form a symmetric matrix). These relations reflect how microscopic time-reversal symmetry affects macroscopic dynamics, and generally no longer holds when time-reversal symmetry is broken. Accounting for applied fields \(B_m\) that break time-reversal leads to the generalised Onsager-Casimir relations, \(L_{jk} (B_m )=L_{kj} (-B_m )\). 
\begin{figure*}
\includegraphics[width=15cm]{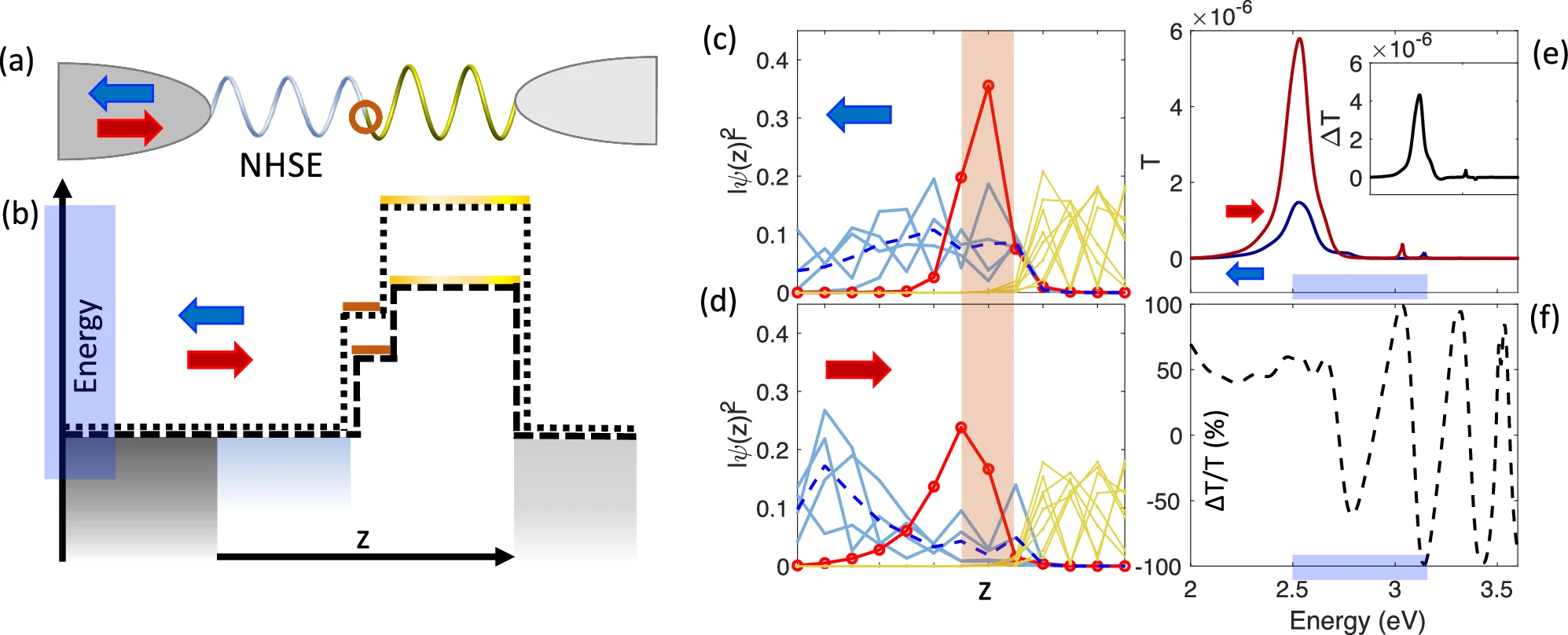}
\caption{Schematic of Zhao \textit{et al.}'s charge-trapping model. (a) the non-Hermitian skin effect gives rise to localisation of interfacial wavefunctions, and interface-magnetisation augments their penetration into the chiral bridge. (b) Impurity/defect-sites act as charge-trapping centers. Localised impurity wavefunctions couple to interfacial wavefunctions, resulting in magnetisation-dependent barrier-potential adjustments, breaking Onsager-Casimir reciprocity and allowing finite anisotropy from the linear-response regime. (Nonlinear-response still dominates in this model due to conductance amplification from barrier-tunnelling.) Crucially, this differentiates biomolecular/semiconductor-CISS from that in chiral metals; free of defect sites, the latter respect Onsager relations. (c) \& (d) Wavefunction spread of the interface (blue), the impurity (red), and the molecular region past the impurity (yellow) across the model for opposing interfacial magnetisations. The red-barred region indicates the impurity-site. (e) \& (f) Transmission probability readings against chemical potential for opposing interfacial magnetisations at zero-bias. Reproduced with permission from Zhao \textit{et al.}\cite{zhao25}, Nat. Commun. \textbf{16}, 37 (2025). Copyright 2025 Authors, licensed under a Creative Commons Attribution (CC BY) license.}
\label{fig:fig7}
\end{figure*}
In the context of two-terminal CISS spin-valve-like set-ups, this means that the charge-conductance measured remains invariant under a reversal of magnetisation (for all magnetised components) and applied magnetic field if present:
$G_{ij} (H,M)=G_{ij} (-H,-M)$. The direct implication is that there should be no anisotropy readings for CISS in the linear regime. Since CISS spin-valve experiments clearly do feature magnetocurrent anisotropy, one may conclude that biomolecular-CISS transport observations are inherently nonlinear whenever such setups do not feature the application of an external magnetic field onto target CISS molecules. This also raises the pertinent question: if SOC is not giving rise to both spin-polarization or conductance anisotropy, then what is being captured in readings? 

This line of thought has given rise to many coarse-grained theoretical analyses, elaborated further in Sec. \ref{part3C}, and it suggests that CISS's anomalous spin-selectivity levels in biomolecular setups may arise from experimental or material-dependent factors not fundamental to CISS (at ambient temperatures), lending well to explicitly nonlinear\cite{diaz_effective_2018} or setup-dependent\cite{alwan_spinterface_2021,dubi_spinterface_2022} accounts. Notably, Dalum and Hedegård \cite{dalum_theory_2019} argue that Onsager-Casimir reciprocity only holds in naïve conditions: spin-orbit coupling establishes a dipole moment across the molecule upon bias application, resulting in altered local-magnetisation for the leads at steady state. In this vein, Liu \textit{et al.} \cite{liu_linear_2020} observe a nontrivial linear contribution to magnetoconductance, attributable to the coercive field of the substrate (Ga,Mn)As giving rise to different conductance states when magnetisation is reversed. Such analyses highlight how the electronic/structural properties of all experimental components are susceptible to change upon bias, pointing towards the importance of nonlinear interactions and feedback between experimental components (see Fig.~\ref{fig:fig6}). A broader introduction to such proposed ‘pictures-in-motion’ is given in Sec. \ref{part3E} on semiclassical models.

\begin{figure*}[t]
\includegraphics[width=12cm]{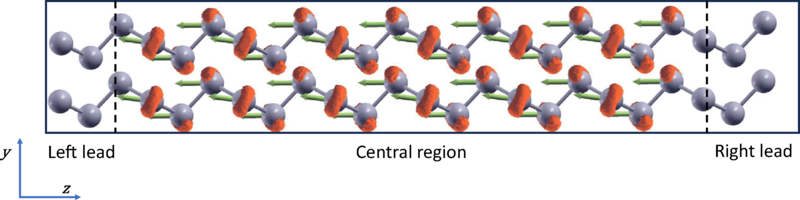}
\caption{Gupta \& Droghetti's (2024) tellurium model schematic. The arrows represent the current-induced magnetic moments, and the orange regions mark the spin-z density isosurface. Time-reversal symmetry is spontaneously broken with bias application. Reprinted with permission from Gupta \& Droghetti \cite{Reena_telluriumTheory_2024}, Physical Review B \textbf{109}, 155141 (2024). Copyright 2024 American Physical Society.}
\label{fig:fig8}
\end{figure*}

Zhao \textit{et al.} \cite{zhao25} elaborate on above by comparing CISS to electrical magnetochiral anisotropy (EMCA), in which an external magnetic field is applied to some chiral conductor; both exhibit non-linear response against a driving electric field and feature induced spin-accumulation in leads like Au. Yet, EMCA preserves Onsager-Casimir relations, leading the authors to propose that biomolecular-CISS’s unique magneto-transport properties may arise from charge-trapping within the dielectric bridge. As outlined in Fig.~\ref{fig:fig7}, due to the non-Hermitian skin-effect, the wavefunction of the magnetised interface may be exponentially localised either towards the interface or spread further into the bridge, depending on magnetisation direction. This interfacial wavefunction couples with localised impurity/defect-sites in the bridge, and altering the wavefunction distribution also modifies the energy of localised impurity states (effectively, the barrier potential), thus changing the electronic properties of the set-up and circumventing the Onsager-Casimir relations \textit{even at zero bias}.

Going beyond anisotropy readings, dominant linear-response in CISS has been amply observed on inorganic chiral media or chiral metals where Onsager-Casimir reciprocity is not expected to fail. (Of note, Yang \& van Wees \cite{yang_linear_2021} suggest that alternative anisotropy measures may instead be used, such as changing the magnitude of magnetisation or a rotation of an applied magnetic field not amounting to reversal.) Transport experiments have, for instance, been conducted without magnetic field application on disilicide crystals\cite{shiota_inorganic_2021} and transition-metal dichalcogenide monolayers\cite{inui_inorganic_2020}, all of which showcase linear current-to-voltage response, attributable to a strong Weyl-type SOC that, given \( \lvert{\alpha_{(x,y)}}\rvert \ll \alpha_{z} \), results in an effective one-particle Hamiltonian:
\begin{equation}
H \cong \frac{\hbar^{2} k_{z}^{2}}{2m} + \alpha k_{z} \sigma_{z} \cong \frac{\hbar^{2}}{2m} \left(k_{z} \pm \frac{m\alpha}{\hbar^2} \right)^{2},   
\end{equation}
which acts as a perturbation to the inversion-breaking crystal field that hybridises the Rashba bands, facilitating spin-momentum locking with monopole-like spin-texture. Most pertinent, however, are helical tellurium nanowires (geometrically akin to \(3_{10}\)-helix peptides); they are naturally \textit{p}-doped semiconductors and can be viewed as direct inorganic analogues of biomolecular junctions in CISS. 

Recent work by Gupta \& Droghetti \cite{Reena_telluriumTheory_2024} employed ab-initio methods to show that for a bundle of tellurium nanowires bonded together by van der Waals forces, collinear magnetic moments emerge upon application of finite bias, spontaneously breaking time-reversal symmetry and giving rise to non-zero spin current in the direction parallel to charge transport within the linear-response regime (see Fig.~\ref{fig:fig8}). It should be noted that nevertheless the spin-resolved conductances remain the same, so their model does not act spin-selectively per se. In such uniform media, as Onsager-Casimir reciprocity remains upheld, magnetocurrent anisotropy becomes an ineffective index. 

\subsection{The Role of Symmetries: Chiral-Induced or Chirality-Induced?}
\label{part2E}

\begin{figure}[b]
\includegraphics[width=8cm]{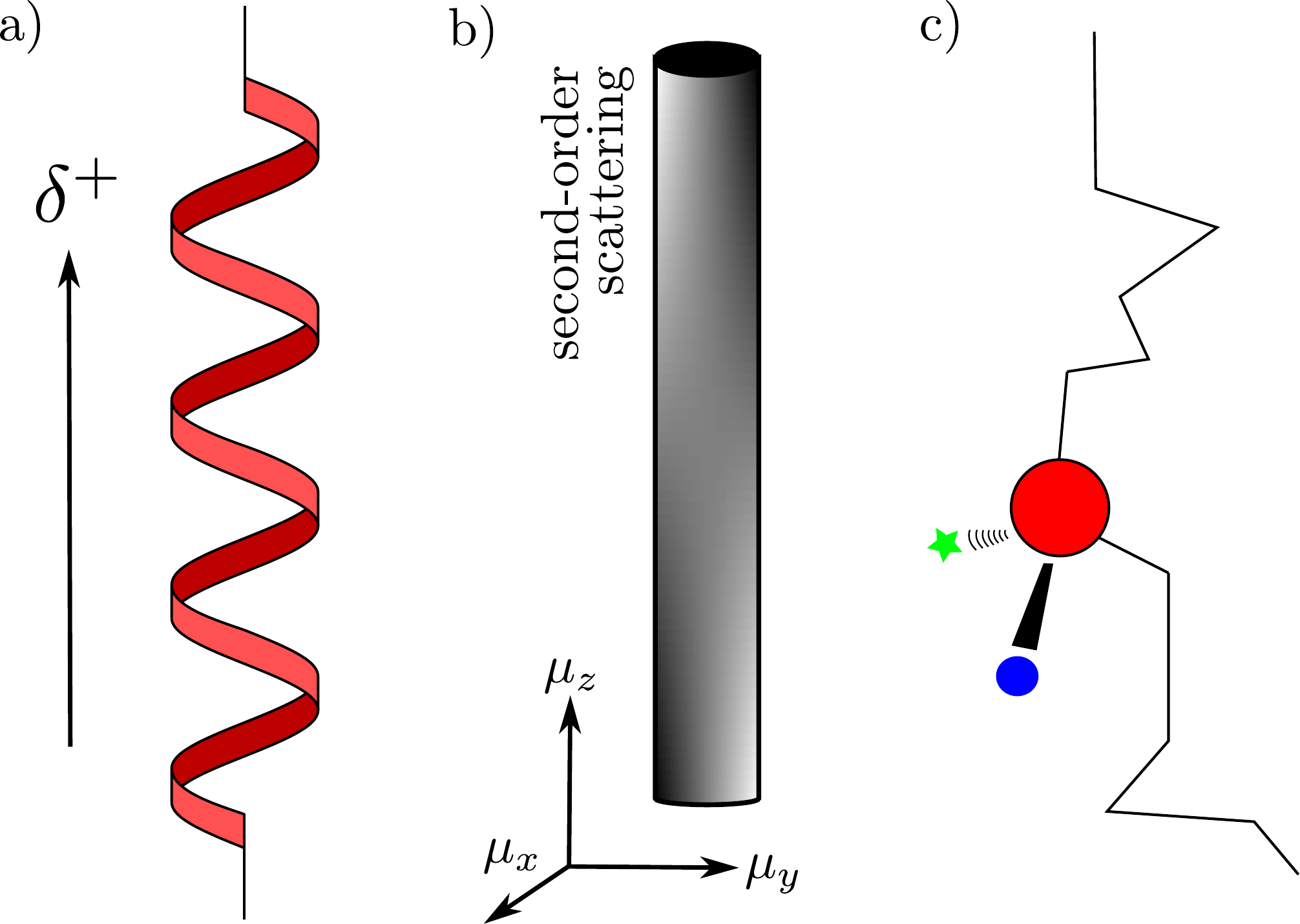}
\caption{Schematic depicting contexts where CISS can be considered: (a) a bridge with helical geometry; (b) a molecular wire equipped with an anisotropic dipole-field with no plainly structural chirality; (c) a molecule with point-chirality but no obvious emergent SOC.}
\label{fig:fig11}
\end{figure}

The prototypical example of CISS occurs within a helical bridge, with non-trivial SOC arising chiefly from a helical geometry. Much of CISS studies revolve around helical structures or molecules. Geometric chirality, however, is broader than helical structures alone, and one example to counter narrow helical-centric interpretations of CISS are the prior-mentioned studies on inorganic chiral crystals, in which inherent chiral magnetic-order/spin-texture spontaneously breaks time-reversal symmetry (we refer to Cheong \& Xu \cite{cheong_magnetic_2022} for an overview). This begs the question, as outlined in Fig.~\ref{fig:fig11}: to what extent is CISS a point-chiral phenomenon as opposed to a helical or geometrically-chiral one? 

The absence of chiral centres \textit{per se} appears irrelevant to CISS insofar as they do not contribute to SOC. Mondal \textit{et al.}\cite{mondal_spin_2021} demonstrated how achiral monomers, when induced to self-assemble into helical macrostructures, produced opposite circular dichroism signals depending on the chirality of the initial solvent. An earlier work by Kulkarni \textit{et al.} \cite{kulkarni_highly_2020} provided evidence of the transport counterpart using magnetic-conductive AFM microscopy, showing that supramolecular helicity was sufficient for CISS. Given this, it is intriguing to note that CISS is not limited to helical molecular structures but encompasses general structural chirality: Ghazaryan \textit{et al.} \cite{ghazaryan_analytic_2020} formulates CISS as a scattering phenomenon under second-order perturbation theory and casts the chiral bridge as an anisotropic dipole-field, retrieving spin-polarization while reproducing length- and energy-dependence trends. Such a dipole-centric perspective has been broached experimentally too, when Eckshtain-Levi \textit{et al.} \cite{eckshtain-levi_cold_2016} observed via non-magnetic Hall measurements that cold-denaturation in alpha-helical oligopeptides can flip spin-polarization direction; simulations suggest that dipole-flips due to rearrangement of dipole-flips upon denaturation may be the underlying mechanism. Such dipole-centric models also implicitly account for charging of the molecular bridge by allowing for in-time non-equilibrium descriptions of CISS, which we discuss further in Sec. \ref{part3E}.

In light of both geometric- and point-chirality contributing to CISS despite the different analyses of their SOC-contributions (the former accountable purely from effective helical-motion, while the latter is better described as emergent from the configuration of $\sigma$- and $\pi$-type orbital overlaps), it is natural to probe at a higher level and question if chirality by itself can give rise to any relevant topological electronic properties \cite{wagner2025probing} in CISS. At its simplest, we can map a 1D helix to a toy Su-Schrieffer-Heeger-like model by assuming continuous screw-sense symmetry and coherent transport. When the sites are arranged such that they admit a perfectly periodic unit-cell, the conservation of screw-sense helical momentum labels the model with a Zak phase that is only quantized when time-reversal symmetry is broken (\textit{e.g.} by an external magnetic field); see Sacramento \& Madeira's \cite{pedro2022} study for a deeper investigation into the topological properties of such 1D chiral electronic chains.  Any spin-selective effect that does not invoke an explicit magnetic field should therefore arise beyond the single Bloch-band description, for instance by employing multi-orbital frameworks to remain within the purely coherent paradigm as demonstrated by Utsumi \textit{et al.} \cite{utsumi_spin_2020, utsumi_2022}. Under the premise that a chiral biomolecule primarily functions as an \textit{orbital-momentum} selector due to its weak SOC, Liu \textit{et al.} go beyond a molecule-only picture and reveal the presence of a topological orbital texture via which orbital-polarisation is generated and then converted to spin-polarisation at the interface with a substrate of non-negligible SOC \cite{liu_chirality-driven_2021}. However, the applicability of either coherent-transport approaches to explain experimental findings is uncertain: initial calculations by Gersten \textit{et al.} \cite{gersten_induced_2013} and recently experimental investigations from Li \textit{et al.} \cite{Li2025_spinRotationExp} using both nonmagnetic and magnetic break-junction setups found that no significant dependence of magnetoconductance or I-V characteristics on enantiomeric handedness within the fully coherent transport regime, owing to interfacial SOC being too weak and transit time being too short for spin-filtering to occur at a coherent time-scale, accentuating the setup-dependent nature for the origin of the CISS effect. More generally, in the absence of external magnetic fields and any crystalline structure, Weyl fermions with finite Chern number appear at time-reversal-invariant momenta for all symmorphic chiral space groups \cite{chang_topological_2018}. Unlike Weyl fermions generated by band inversion, these chiral fermions exist \textit{a priori} and their finite chiral charge implies that they can only be destroyed via pairwise annihilation, requiring application of a large magnetic field. Whether they manifest in chiral soft-matter is an intriguing avenue for further work, but it suffices to note at current that chirality is not a sufficient factor for spin-selectivity.

\subsection{That Which We Call CISS: Open Questions and Avenues}
\label{part2F}

What is CISS? The breadth of CISS belies the difficulty of building an overarching framework capable of encompassing all its realisations. On one hand, CISS occurs in deformable dielectric helical biomolecules/films in noisy ambient conditions, exhibiting nonlinear transport with strong substrate effects. On the other, CISS occurs in non-centrosymmetric crystalline structures and nanowires with none of the complications the former case brings. Both directions are equally rich with promise: the former explores spin effects in biology and the rise of biomimetic quantum/spintronics technologies, while the latter can open the door to unconventional superconductivity \cite{PavanHosur_superconductor_2014} and other magnetochiral effects. Controversy within the theoretical literature is symptomatic of this split in the field, and caution should be taken by new entrants therefore to ground their investigations in CISS within the assumptions of their respective fields. We argue that CISS can be briskly summarised under the following checklist:
\begin{enumerate}
    \item Geometric inversion-symmetry breaking: A noncentrosymmetric structure or helical geometry gives rise to a spin-orbit coupling field. 
    \item Rashba-likeness: Such an intrinsic spin-orbit coupling field gives rise to a magnetic order, and thereby a spin-texture and spin current.
    \item Selectivity: The presence of a spin current does not guarantee spin-selectivity or preferential transmission of spin-species; that lies in the purview of the retrieval or conversion mechanism within the physical setup.
    \item Confounders/Enhancers: Electromagnetic properties of chiral object(s) can be altered by external setup components and vice-versa, potentially leading to anomalous effective SOC strengths and nonlinear response.
\end{enumerate}

Beyond controversy in the traditional realisations of CISS-transport and CISS-transmission, novel observations on CISS have lead to surprising developments in both theory and application. One such example is the potential use of CISS for orbitronics or orbital-selectivity mechanisms\cite{wolf_unusual_2022, adhikari_interplay_2023, liu_magnetless_2024, liu_chirality-driven_2021}, especially concerning media with negligible SOC. Additionally, non-adiabatic corrections for nonequilibrium nuclear motion can strongly affect reaction evolution and also transport characteristics in molecular junctions \cite{kershaw_nonadiabatic_transport_2017, kershaw_nonadiabatic_transport_2020}, leaving the extent of Berry effects in the neighborhood of (Kramers-)Weyl fermions an open question in CISS \cite{bian_modeling_2021, bian_meaning_2022}. And while the theoretical disagreements around CISS appear to arise largely from contextual differences, many open questions still remain as to the broader implications in emerging quantum engineering prospects for CISS. For example, do thermal fluctuations and phononic vibrations enhance or stymie CISS? Does CISS play a role in chiral transmission? Where does CISS fit into the broader realm of chiral phenomena from magnetic chirality to chiral quantum optics \cite{cqo25}? Most recently, Briggeman \textit{et al.} exhibited analog simulation of 1D chiral molecular conductors by using AFM lithography to etch planar channels on an oxide-interface that explicitly break mirror symmetry in the electron potential landscape, thereby offering direct means to isolate purely geometrical chiral SOC effects \cite{Megan2025_analogSimul}. In concluding this section, we hope that our discussion helps dispel some incertitude for those new to the area pertaining to the contentiously fractured state of the wider CISS literature, and we will now follow in the next section with a broad overview of theoretical efforts in CISS.

\section{Within the Gap: Models, Mechanisms, And The Missing Link}

Having surveyed some foundational issues underpinning CISS theory, we now turn to organizing the key modeling approaches that have been advanced to explain the CISS effect within each broad class of experimental manifestations, running through excited-state frameworks to continuum models and tight-binding treatments before ending in spin chemistry and electron-transfer approaches to CISS. Through this, we hope to highlight the different regimes these separate modeling approaches cover, thus laying the conceptual ground to briefly discuss emerging applications, such as functional molecular devices, next-generation LEDs, plus quantum sensing and detection platforms, which we address in Sec. \ref{part4}. 

\subsection{CISS in Excited-States: Theory behind Optical-Activity in Chiral Molecules under Photoexcitation}
\label{part3A}

In Ray \textit{et al.}’s \cite{ray_asymmetric_1999} initial paper, photoelectrons emitted from L-/R-stearoyl lysine biofilms exhibited asymmetric transmission using clockwise circularly-polarized light over its counter-clockwise counterpart, marking the first reported observation of CISS. Subsequently, Göhler \textit{et al.} \cite{gohler_spin_2011} found using Mott polarimetry that a chiral ds-DNA film bound to Au displayed spin polarization even with linearly-polarized light, confirming chirality can by some mechanism induce spin-selectivity in transmission. But to what extent are these results attributable directly to transmission across the chiral film than they are indirectly to interfacial effects of chiral-film-adsorption upon the substrate? To better understand photoemission in CISS, it is useful to go over how natural optical activity arises in chiral bodies. This is well-covered by the theory of circular dichroism, in which the differential absorption of clockwise and counterclockwise circularly-polarized light is measured \cite{katsantonis23}. Of note, a concise treatment for the lone electrically conducting helical-molecule case appears in Andrews \& Tretton \cite{andrews_physical_2020}, providing an intuitive classical analogy.

Heuristically, the rotating electric field of circularly-polarized light induces a static dipole moment along its direction of propagation, potentially altering electronic properties via mobile charge/spin-rearrangement and affecting the molecules’ absorption spectrum; and the rotating magnetic field acts as an effective Zeeman-splitting field for electrons along the longitudinal (if perpendicular to the helical axis) or transversal (if parallel) directions, assuming the fields do not average to zero. However, gaseous/liquid-state chiral bodies (for which a lone-molecule approximation would be justified) are seldom used to probe CISS-transmission due to random dispersion/orientation of samples. Rather, CISS is tested on ferromagnetic substrates functionalised with organic chiral films. As such, photoelectrons are collected from both the chiral film and the substrate surface, with spin-polarization arising from both components and their interplay.
\begin{figure}
    \centering
    \includegraphics[width=1\linewidth]{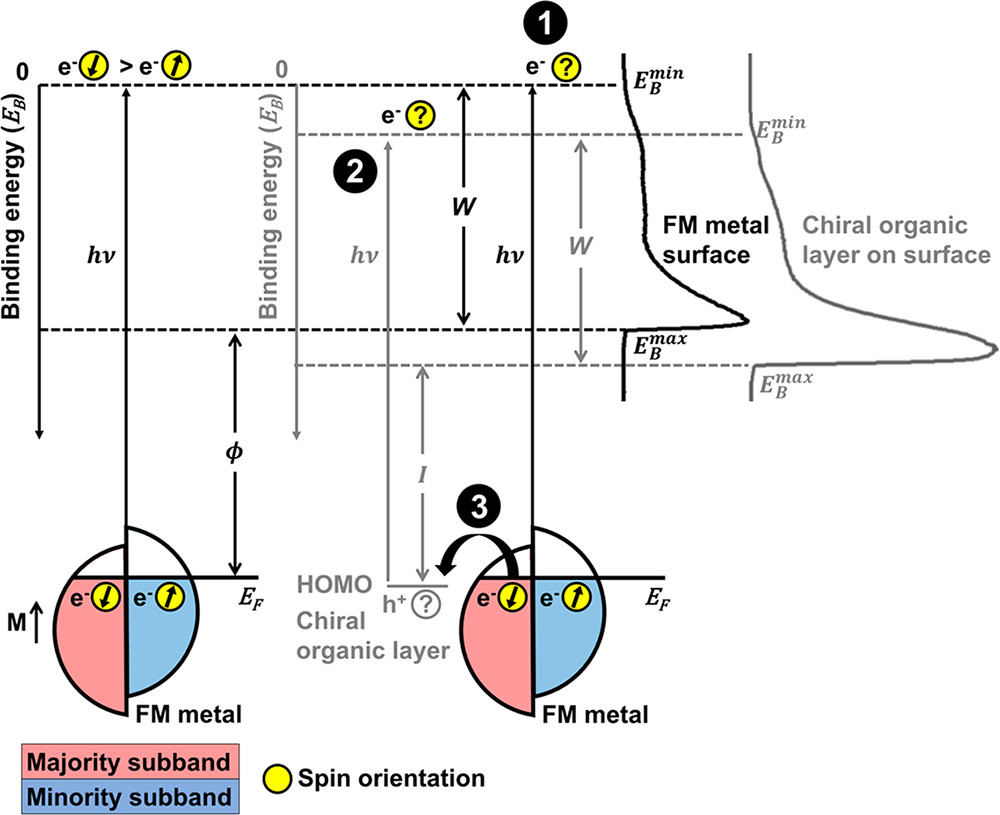}
    \caption{Schematic by Abendroth \textit{et al.} illustrating photoelectron spectroscopy on ferromagnet-substrate-chiral film setups and representative ultraviolet photoelectron spectra, annotated with routes to spin-selective emission: ﬁltering of photoelectrons originating from the ferromagnet by the chiral thin-film (1), spin-polarized photoemission, indirectly inﬂuencing photoionisation energy (I) of the chiral-ﬁlm (2), or ﬁltering of conduction electrons supplied by the metal to ﬁll holes left in the valence orbitals of the organic ﬁlms, represented as highest-occupied molecular orbitals (HOMO) (3). In the case of linearly polarized light being incident on the setup, route (3) can still generate spin-selection. Reprinted with permission from Abendroth \textit{et al.}\cite{abendroth_spin-dependent_2019}, Journal of the American Chemical Society \textbf{141}, 3863-3874 (2019). Copyright 2019 American Chemical Society.}
    \label{fig:fig12}
\end{figure}
Abendroth \textit{et al.} \cite{abendroth_spin-dependent_2019}, in an experimental study on the valence electronic structure of CISS-FM layers, propose that photoelectrons undergo spin-selectivity under three routes: firstly, scattering of FM surface electrons through the chiral fields (enhanced by proximity effects of the packed layer) experiences a lower work-function if the majority-spin state in the FM is also preferred by the chiral layer; secondly, incident photons can ionize adsorbed chiral-molecules given a high enough energy and emit photoelectrons from the chiral layer (thus forming a main source of spin-polarized photoemission differentiable between oppositely circularly-polarized light); thirdly, photoelectrons from the FM may collide with and ionise chiral molecules, resulting in secondary scattering processes; and fourthly, as the chiral layer builds up positive charge due to photoionisation, conduction electrons are induced to transfer into valence holes in the chiral layer, as seen in Fig.~\ref{fig:fig12}. This process is further compounded by spin-transfer torque effects when a magnetic order is induced in the substrate upon adsorption of a chiral layer \cite{ben_dor_magnetization_2017}.

\subsection{CISS in Excited-States: Scattering Models for Photoexcitation}
\label{part3B}

Given the complications presented above, a prominent goal has been to find some analytical description that captures the essential physics of CISS-transmission. A foundational effort is given by Yeganeh \textit{et al.} \cite{yeganeh_chiral_2009}, in which scattering theory is used to model electron transmission through a helical potential dressed with a spin-orbit coupling field, assuming elastic first-order scattering.  Their results yield that a finite pitch result in differential angular momentum transfer, validating chirality’s role in spin-selectivity. However, asymmetry readings are too low to match experimental results, rendering chirality alone to be a necessary but insufficient factor.
% \begin{figure}
%     \centering
%     \includegraphics[width=0.5\linewidth]{figs/sourced/yeganeh.jpeg}
%     \caption{Single-scattering model used by Yeganeh \textit{et al.}, detailing the scattering angle ($\theta$) and the momenta before/after the scattering ($\mathbf{k}_a$/$\mathbf{k}_b$), while $\varphi$ parameterizes the helix with a single path variable.}
%     \label{fig:fig13}
% \end{figure}
Subsequent work expands on this approach by considering the multiple-scattering regime \cite{medina_chiral_2012}, interfacial conversion of orbital-to-spin angular momentum due to substrate SOC \cite{gersten_induced_2013}, inelastic effects \cite{varela_inelastic_2013}, and a minimal model that generalizes the CISS-helix into a CISS-dipole \cite{ghazaryan_analytic_2020}. However, these analytic methods either do not give rise to anomalous spin-polarization readings as measured in Mott polarimetry, or cannot (in the case of interfacial effects) account for similar spin-polarization readings when substrates with low-SOC are used \cite{kettner_chirality-dependent_2018}. It stands to reason that either some non-equilibrium \cite{pbs25}mechanism perturbs the set-up significantly that is not captured by the above methods, or there is some omitted factor in the chiral molecule itself that is responsible. Could there be more to the helical contribution to CISS than currently treated? We now move to discussing CISS-transport in waveguides in the next subsection, where we cover efforts to find some rigorous starting-point to study CISS. 

\subsection{CISS in Waveguides: Continuum Approximations and Geometric Efforts}
\label{part3C}

That helicity itself is a main contributor to spin-selectivity is an attractive idea at first glance, but how exactly can helical contributions be quantified? A continuum formulation of CISS, though rarely featured in the literature, can serve as a robust starting-point to derive an effective low-dimensional Hamiltonian and recover geometric effects. One convenient method to study helicity is to employ a rotating Frenet-Serret frame, allowing the curved path to be parametrized by a single path variable while the curve’s torsion and curvature then determine the frame’s rotation. Infinite potential walls (implying adiabatic separation of tangential and transversal Hamiltonian components) can then be imposed along this path according to da Costa’s thin-layer quantization method \cite{da_costa_quantum_1981}.

As per da Costa, these non-inertial effects are captured via a ‘geometric potential’ that can be appended to the Pauli-Schrodinger Hamiltonian. Several early works \cite{gutierrez_modeling_2013} start from this point but do not explicitly investigate curvature-effects. Michaeli \& Naaman \cite{michaeli_origin_2019} begin by considering a tubular waveguide rather than a quasi-1D wire, adding a strong dipole potential $V_D$ growing linearly along the molecular axis. Likening CISS to a Klein-tunneling phenomenon, they describe their system with the Hamiltonian:
\begin{equation}
    H_{N,l}(s) = E_N + V_D(s) + \frac{\mathbf{p}_s^2}{2m} + \frac{\mathbf{p}_s}{m} \cdot \hbar \gamma l + \kappa \vec{\sigma}\cdot\vec{L}_{helix} + \Delta E_l,
\end{equation}
whose eigenstates are a product of two wavefunctions, one a harmonic oscillator in the plane perpendicular to the helical axis ($s$) and the other dependent on $s$, such that $E_N$ refers to the energy level of the $N$-th state of the former, $V_D$ the potential along $s$, the third-last term a centripetal-like potential, the second-last term SOC due to the confining helical potential, and the final term being a curvature-dependent energy-shift. And after applying the gauge transformation,  $\ket{\sigma_z^{\pm}} \rightarrow \exp[\pm(i\pi s/R)\cdot \ket{\sigma_z^{\pm}}]$, they 
observe the emergence of a momentum-independent Zeeman-like component, $\kappa l \sigma_y $, from the rotating atomic SOC field. Here, the strong dipole component is proposed to extend CISS’s energy range up to that of the dipole potential while enhancing spin-polarization by suppressing transmission of unfavorable spin-states, using Airy functions to model wavefunction decay. It should be duly noted that though the appearance of a Zeeman-like field brings about finite asymmetry in spin-transmission, this model does not break time-reversal symmetry at the Hamiltonian level. Geyer \textit{et al.} \cite{geyer_effective_2020} then build on this work, investigating the validity of applying the thin-layer quantization method under a generic (scalar) SOC field. They demonstrate that the momentum-independent Zeeman-like term, given by $vl\vec{\sigma}\cdot\vec{B}$ in their derivation, arises from transversal degrees of freedom, originally omitted from Gutierrez \textit{et al.}’s \cite{gutierrez_modeling_2013} treatment, preventing removal of SOC via gauge transformation.

However, these efforts take the Pauli-Schrodinger equation as their starting point which describe a Dirac fermion in an inertial frame, which the Frenet-Serret frames are not. Shitade \& Minamitani \cite{shitade_geometric_2020} attempt to circumvent this by employing the generally-covariant Dirac equation, generating the Langragian directly from holonomic constraints to derive an appropriate single-electron Hamiltonian:
\begin{equation}
    H = \frac{\mathbf{p}_s^2}{2m} + \frac{\hbar^2 \kappa^2}{8m} + \frac{\hbar}{2}\cdot \frac{\{\mathbf{p}_s, \kappa\vec{\sigma}\cdot\vec{B}\}}{2},
\end{equation}
after applying the Foldy-Wouthuysen (non-relativistic) and the thin-layer approximations. Difficulty arises in this approach, however, as changing the order of approximations applied alters the form of the Hamiltonian, reflecting a necessity for further work into canonical quantization. Alternative curvature-focused continuum-model efforts that do not make use of thin-layer quantization do exist but are uncommon, like the scattering model of Medina \textit{et al.} \cite{medina_chiral_2012}, or Diaz \textit{et al.} \cite{diaz_effective_2018} where molecular deformation is accounted for by deriving an effective nonlinear model for electron transport \cite{qTr25} through a helix composed of vibrating point dipoles, simplifying CISS to a Davydov-like model.

One conclusion common to efforts in this vein is that curvature-effects give rise to an effective momentum-independent Zeeman-splitting field. This appears to contradict Kramers' degeneracy, in which time-reversal symmetric systems should display equal transmission eigenvalues for both spin-configurations \cite{Bardarson_2008}. Furthermore, Gersten \textit{et al.} \cite{gersten_induced_2013} demonstrated heuristically that on regular Cartesian coordinates, spin-flip contributions in the single-scattering regime and spin-precession throughout helical motion are both negligible, with spin-flip transition probability to order $\sim10^{-9}$ and an estimated spin-rotation length-scale of $\sim10^4$ relative to helical pitch. Such results are in line with Mashhoon's \cite{PhysRevLett.61.2639} spin-rotation coupling term, given by $\delta H_{SR} = \gamma(\delta H_{Thomas} - \boldsymbol{\Omega} \cdot \mathbf{S})$, which is vanishingly small at below-relativistic speeds.

\begin{figure}
    \centering
    \includegraphics[width=0.9\linewidth]{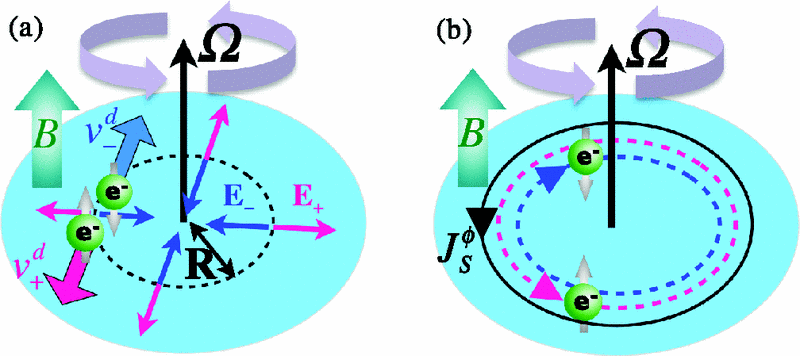}
    \caption{Illustration by Matsuo \textit{et al.} of their spin-mechatronics model. When a plane is made to rotate at some angular velocity and an external magnetic field is applied normal to the plane, opposite spins experience opposite effective effective fields, giving rise to opposite spin-currents. This can be interpreted as an inverse Einstein-de Haas effect. Reprinted with permission from Matsuo \textit{et al.}\cite{matsuo_current_2011}, Physical Review B \textbf{84}, 104410 (2011). Copyright 2011 American Physical Society.}
    \label{fig:fig14}
\end{figure}

Though questions regarding quantization methods remain at-large, field-theoretic approaches may still be adopted. For instance, Matsuo \textit{et al.}’s \cite{matsuo_effects_2011, matsuo_current_2011} procedure to generalise the Pauli equation for non-inertial frames can be used to provide a rigorous background to Gersten \textit{et al.}'s heuristics. Since Gersten \textit{et al.} assume a constant $z$-velocity for the electron, we can also model the electron as a particle on a rigid rotating rod on an inertial plane that is moving at some constant velocity in the $z$-axis with respect to an observer (Fig.~\ref{fig:fig14}). This allows us to study non-inertial effects of a particle on a constrained helical path in regular Cartesian coordinates, making it tractable to comparison while bypassing the quantization procedure, resulting in the single 2-spinor free Hamiltonian: 
\begin{equation}
\begin{split}
    H &= \frac{\vec{\pi}^2}{2m} + qA_0 + \mu_B \vec{\sigma}\cdot\vec{B} - \vec{\Omega}\cdot(\vec{r}\times\vec{\pi} + \frac{\hbar}{2}\vec{\sigma})\\ &+ \frac{\lambda}{2\hbar} \vec{\sigma} \cdot [\vec{\pi} \times q\vec{E}' - q\vec{E}' \times \vec{\pi}] -\frac{\lambda}{2}\mathrm{div}[q\vec{E}'],
\end{split}
\end{equation}
where the electric field is augmented as $\vec{E}'=\vec{E} + (\vec{\Omega}\times\vec{r})\times\vec{B}$, accounting for a Lorentz boost due to an external magnetic field. Note that the $\lambda$ denotes $\hbar^2 / (4 m_0^2 c^2)$ but is used more generally in the literature to refer to phenomenological tight-binding SOC strength. Since SOC is to order $m^{-2}$ while the spin-rotating term is to order $m^{-1}$, SOC contributions can be considered small in vacuum conditions; and in the absence of an external magnetic field, the spin-rotation coupling, $\Omega \cdot \sigma$, provides spin-splitting. Though this analysis corroborates the existence of geometric effects in CISS, their strength remains a matter of contention. On one hand, heuristically accounting for a linear electronic momentum of 1eV as with Gersten \textit{et al.} \cite{gersten_induced_2013} makes such effects negligible. On the other hand, recent work from Bradbury \textit{et al.} \cite{bradbury2025} account for non-Born-Oppenheimer effects using a phase-space formulation of electronic structure, predicting a strong spin-Coriolis effect even for electrons of small molecules. And while there are innovative schemes to explicitly test curvature-dependence of CISS in experiments\cite{shekhter_magnetoconductance_2022}, much of biomolecular-CISS as observed is a non-equilibrium phenomenon where the effects of electron correlations and electron structure in materials cannot be neglected. Accounting for these likely-more-significant contributions calls for a different approach.

\subsection{CISS in Waveguides: Coarse-Grained Models}
\label{part3D}

\begin{figure}
    \centering
    \includegraphics[width=0.7\linewidth]{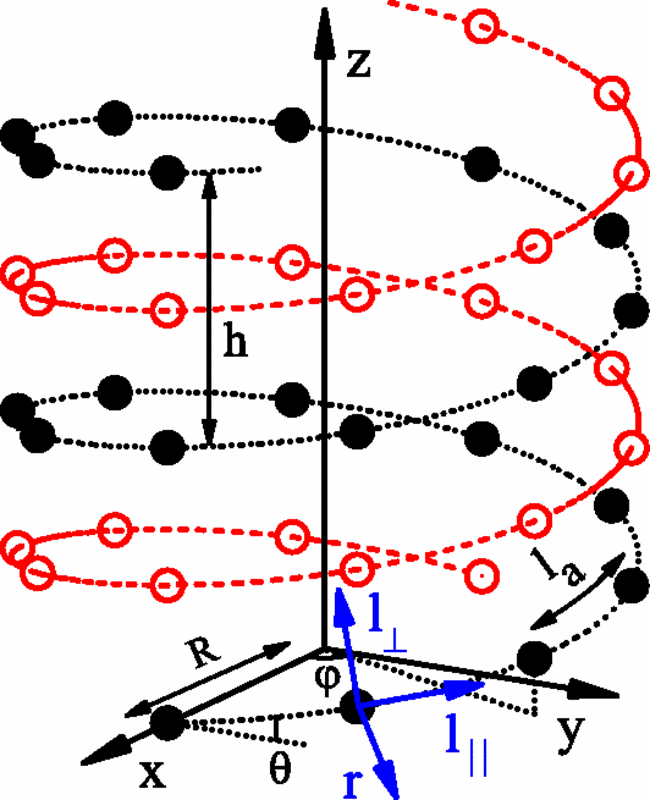}
    \caption{Guo \& Sun's (2012a) tight-binding model of CISS, where each atom is modeled by a node fitted with its own Frenet-Serret frame, parametrized by a path variable ($\varphi$). Each Frenet-Serret frame is oriented such that the tangent unit-vector is tangential to the helix. Reprinted with permission from Guo \& Sun \cite{guo_spin-selective_2012}, Physical Review Letters \textbf{108}, 218102 (2012). Copyright 2012 American Physical Society.}
    \label{fig:fig15}
\end{figure}

While continuum models are useful in distilling geometric contributions to CISS and scattering effects, electrons do not simply tunnel through free-space nor along helical waveguides. For non-excited state transfer, transmission is mediated via both through-space and through-bond couplings, the latter of which arises from orbital interactions.  From this, two general mechanisms of electron transport may feature: (a) a coherent, single-step tunnelling process, predominant at short distances/lengths; (b) an incoherent, thermally-activated multi-step hopping scheme; alongside intermediate mechanisms (structural fluctuation-induced ‘flickering resonance’ etc.). Though schema have been devised to determine which mechanisms dominate in an experimental set-up\cite{nurenberg_evaluation_2019}, it would be desirable to arrive at numerical transport predictions. Here, a coarse-grained approach employing tight-binding models captures core elements of actual setups while keeping flexibility and tractability in adjusting parameters. Prototypical efforts in this vein were given by Gutierrez \textit{et al.}\cite{gutierrez_spin-selective_2012} and Guo \& Sun\cite{guo_spin-selective_2012}; both of which made use of simplified single-orbital/channel tight-binding models, featuring parameterized coupling strengths and variable SOC schemes (Fig.~\ref{fig:fig15}). The leads can be accounted for explicitly, and transmission spectra and a projected local density-of-states of the entire apparatus can be treated by the Landauer-Buttiker formalism, or non-equilibrium Greens' function methods to account for dissipative effects.
\begin{table}
    \centering
    \begin{tabular}{|c|c|}
        \hline
        Description & Expression \\
        \hline
        Kane-Mele-like & $\sum_{m} i\lambda \psi_m^{\dagger} (\vec{d}_{m,m+1} \times \vec{d}_{m+1,m+2}) \cdot \mathbf{\sigma} \psi_{m+2} + \mathrm{H.c.} $ \\
        \hline
        Rashba-like & $\sum_{m} i\lambda \psi_m^{\dagger} (\sigma_m + \sigma_{m+1}) \psi_{m+1} + \mathrm{H.c.}$ \\
        \hline
        Aharonov-Casher & $t\sum_m \psi_m^{\dagger} \psi_{m+1} + \mathrm{H.c.} \longrightarrow$ \\
        & $t\sum_m \psi_m^{\dagger} e^{i\lambda(\vec{d}_{m,m+1} \times \mathbf{E}_{m,m+1})\cdot\sigma} \psi_{m+1} + \mathrm{H.c.}$\\
        \hline
    \end{tabular}
    \caption{Phenomenological SOC expressions. $\lambda$ denotes the phenomenological SOC strength; $\vec{d}_{m,n}$ denotes the bond vector from site m to n, while $E_{m,n}$ denotes the local electric field on that bond.}
    \label{tab:table1}
\end{table}
Of particular note, the literature features a broad array of SOC quantization schemes, each reflecting the background under which CISS is being treated. A number of such bare-SOC descriptions are stated in Table \ref{tab:table1}, each corresponding to a subtly different physical picture, making comparison between models somewhat tenuous: the Kane-Mele-like expression considers intrinsic SOC in which inversion-symmetry negates nearest-neighbor SOC coupling and leaves only next-nearest-neighbor (NNN) SOC terms surviving; the Rashba-like expression instead assumes the Frenet-Serret frame and computes a nearest-neighbor SOC component between the rotated spinors; and in a similar vein, the Aharonov-Casher-like phase description is motivated by the fact that an applied electric field $F(t)$ is proportional to the time
derivative of the time-dependent vector potential $\mathcal{A}(t)$ via $F(t) = -c^{-1}\partial_t \mathcal{A}(t)$, such it augments the nearest-neighbor hopping term by a factor $\exp[i\gamma \mathcal{A}(t)]$ and explicitly breaks time-reversal symmetry. 

It can be argued that choice of (bare-)SOC scheme does not affect the overall physics of a CISS model given the low orders of SOC coupling, though a symmetry analysis from Kohn-Sham DFT calculations by Z\"ollner \textit{et al.}\cite{zollner_influence_2020} suggests that imaginary components of an SOC field lead to nonsymmetric Green’s functions/propagators that give rise to finite spin-current. Regardless of the uncertainty on SOC and geometric contributions, it is generally agreed that bare-SOC cannot make up for the anomalous extent of spin-selectivity observed in the ambient noisy conditions CISS remains stable under. Much of the theoretical literature revolves around the search for the missing link to bridge CISS’s strength gap alongside other key features, with coarse-grained models covering a vibrant exchange of ideas owing to the convenience the method affords. Early efforts from Gutierrez \textit{et al.} \cite{gutierrez_spin-selective_2012} to build a minimal model for CISS use a Peierls-substitution SOC scheme to suggest that despite low SOC strengths, small charge-mobility (due to weak dispersion) enhances the extent of spin-flip/rotation in a general model, benchmarking spin-polarization to average spin-transmission asymmetry. This is immediately challenged by Guo \& Sun \cite{guo_spin-selective_2012} who, featuring a second-quantized SOC scheme on a rotating frame, surmise that, for dissipationless single-channel models with ideal unpolarized leads, the SOC field can be removed by a unitary transformation that recasts the system in a rotating frame following spin precession (additionally, CISS was benchmarked here using charge-conductance anisotropy). 
The contrast between these two works sets some basic conditions around which further coarse-grained models build:
\begin{itemize}
    \item for single-channel models, dephasing \cite{dephasing23} or decoherence (arising from electron-phonon or electron-electron interactions, etc.) is necessary to break Kramers’ degeneracy produce spin-selectivity;
    \item time-reversal symmetry on the Hamiltonian level is assumed to hold under a lack of external field application;
    \item while the leads may be considered unpolarized to isolate helical contributions to spin-selectivity, this is only justified if non-linear response sources (i.e. bridge polarizability, charging effects, electron-electron correlation) and non-equilibrium effects such as electron-reorganization or bridge-deformation can be neglected, which may no longer correspond to realistic pictures;
    \item either spin-resolved transmission or spin-current anisotropy or charge-current anisotropy is used to benchmark spin-selectivity in CISS models, none of which are equivalent to each other.
\end{itemize}
In effect, a coarse-grained approach lends well to studying idealized scenarios or a `part-by-part' investigation of CISS. Using parameterized coupling strengths derived from experimental results and ab initio calculations, Guo \& Sun explored physically-motivated double-stranded (dsDNA: 2012a,b) \cite{guo_spin-selective_2012,guo_sequence-dependent_2012} and single-stranded models (polyalanine: 2014) \cite{guo_spin-dependent_2014}. However, given bare-SOC’s inability to account for CISS’s anomalous spin-selectivity, future analytic phenomological models have since shifted to investigating model additions or specific interaction contributions. To summarize explorations within this line:\\

\begin{itemize}
    \item multi-terminal effects on single-channel models have been found to cause dephasing via interference like Buttiker virtual leads \cite{guo_spin-dependent_2014, matityahu_spin_2017};
    \item modeling transport across multiple orbitals (often the three p-orbitals) allows for finite spin-selectivity with fully-coherent transport \cite{utsumi_spin_2020} since SOC can no longer be removed by a gauge transformation. This has also been demonstrated using explicit Slater-Koster formulation of orbital-overlaps, accounting for the exact paths by which SOC emerges and deriving effective couplings \cite{varela_effective_2016, geyer_chirality-induced_2019}; alternatively, a large effective SOC may arise from spontaneous formation of electron-hole pairs between s- and p-orbitals \cite{xiaopeng_spontaneous_order_2020}. 
    \item electron-lattice interactions can greatly enhance spin-selectivity owing to lower charge-mobility. Electron-phonon interactions in particular can induce indirect electron-electron attraction, leading to exchange splitting across spin channels that culminates in a temperature-activated/vibrationally-assisted spin-selectivity  \cite{fransson_vibrational_2020, fransson_charge_2021, sang_temperature_2021, das_temperature-dependent_2022, vittmann_delocalised_phonons_2022}. Such electron-lattice coupling can also be represented by non-adiabatic coupling to nuclei that manifest as a finite Berry curvature \cite{teh_spin_2022, bian_modeling_2021, bian_meaning_2022}, or phenomenologically described by a dissipative frictional non-linear potential  \cite{volosniev_interplay_2021}.
    \item polaron formation (where applicable from strong coupling between electron and lattice) is proposed to induce spin-momentum locking in which an “inverse Faraday effect” favors spin-flips and suppresses backscattering due to the relatively large mass of polarons, indirectly enhancing spin coherence  \cite{zhang_chiral-induced_2020};
    \item similarly, electron-electron correlations (Coulomb interactions) can also enhance spin-selectivity by inducing spin-asymmetric spectral weights while also being sensitive to external magnetic conditions such as lead polarization  \cite{fransson_chirality-induced_2019, dianat_role_2020, kapon_probing_2023};
    \item non-equilibrium conditions (i.e. presence of a bias field) are necessary for CISS as detailed by Wolf \textit{et al.} \cite{wolf_unusual_2022}, unless an external magnetic flux is applied \cite{chensong_persistent_2024}. Additionally, Fransson \cite{fransson_chirality-induced_2019, fransson_vibrational_2020, fransson_charge_2022} argues strongly and are the first to account for the presence of ferromagnetic or polarized-leads. Dalum \& Hedegard \cite{dalum_theory_2019} go further to explicitly demonstrate that CISS alters magnetization/polarization in the leads due to electron-reorganization.
    \item many miscellaneous studies including: contact/lead chirality effects \cite{dednam_group_theory_2023}; gate voltage effects\cite{pan_effect_2015}; spin-current-induced mechanical torque\cite{sasao_spin-current_2019}; spin-current-induced charge-current\cite{fransson_2024, shiota_inorganic_2021}; chiral phonon effects \cite{fransson_chiral_2022, chen_chiral_2022} (see Wang \textit{et al.}, 2024\cite{Wang_2024_chiralphonons} for a broad overview); waveguide-transfer effects \cite{vittmann_interface-induced_2022}; Floquet-driving effects \cite{Lopez_2022, phuc_2023floquet}; realization of Majorana modes \cite{chensong_majorana_2024}; an explicit Lindbladian treatment with dephasing \cite{guosun_2025}; et cetera.
\end{itemize}

Aside from a handful of works that treat CISS as a fundamental mirror-symmetry breaking phenomenon in search for a general theory closer to application in chiral crystals or inorganic chiral nanowires, many efforts strive to account for CISS as observed in biomolecular experiments. In this regard, tight-binding models are ill-at-ease with deformable biomolecules and similar dielectric soft matter. \textit{ab initio} numerical techniques present a powerful avenue in accounting for these details and have already been employed to study symmetry or conformational effects \cite{zollner_insight_2020, zollner_influence_2020}, interface effects \cite{naskar_2023}, and the role of many-body exchange interactions \cite{dianat_role_2020} in CISS, with Dednam \textit{et al.}\cite{dednam_group_theory_2023} showing that achiral molecules can give rise to spin-polarization if the leads/contacts are chiral.

Additionally, there has also been a vigorous thrust to reconsider CISS’s identity as a spin-selectivity phenomenon and broaden it to include orbital-selectivity due to spin-orbital mixing by SOC. This line of argument stems back to the earliest theoretical work on CISS by Skourtis \textit{et al.}\cite{skourtis_chiral_2008}, in which they propose that enhancement of CISS-transmission via circularly-polarized light of the matching helicity operates via ‘current transfer’: or the (partial) conservation of angular-momentum as electrons transfer from a donor to a bridge. Following up on this, Gersten \textit{et al.}\cite{gersten_induced_2013} analytically study how spin-selectivity in CISS arises from interfacial spin-orbit coupling and orbital angular-momentum selection by the molecular bridge’s helical geometry, showing that they eliminate the need for dephasing or inelastic effects. This culminates into Liu \textit{et al.}’s \cite{liu_chirality-driven_2021} argument that in CISS, the chiral molecule acts primarily to select orbital angular momentum, inducing a topological orbital texture that is reliant on inversion-symmetry breaking rather than strict helical geometry; spin-selectivity is conferred by SOC in the leads or at the interface, forgoing reliance on the low SOC strength of chiral biomolecules altogether. Experimental encouragement for this viewpoint has been found in the efforts of Adhikari \textit{et al.} \cite{adhikari_interplay_2023}, where use of a low-SOC substrate (Al) did not generate CISS in MR-response as compared to Au. While the microscopic picture of how a chiral, not necessarily helical, geometry influences spin-dynamics remains unclear, there are strong signs that the orbital degree-of-freedom cannot be neglected and that CISS should be viewed as a generalized angular-momentum phenomenon. This detour highlights the importance of whole-setup effects, which we explore in more detail in the next section.

\subsection{CISS-on-Substrates: Pictures-in-Motion and Extra-Molecular Effects}
\label{part3E}

While coarse-grained models are instructive in determining the necessary conditions to and contributors for CISS, most of them fall short in accounting for its non-equilibrium, non-linear behavior, compounded by whole set-up effects that cannot be ignored in trying to account for CISS’s strength. In the absence of powerful ab-initio methods that can simulate entire CISS set-ups, what is necessary to better understand non-equilibrium effects in CISS and distill their contributions are time-dependent models.
\begin{figure}
    \centering
    \includegraphics[width=1\linewidth]{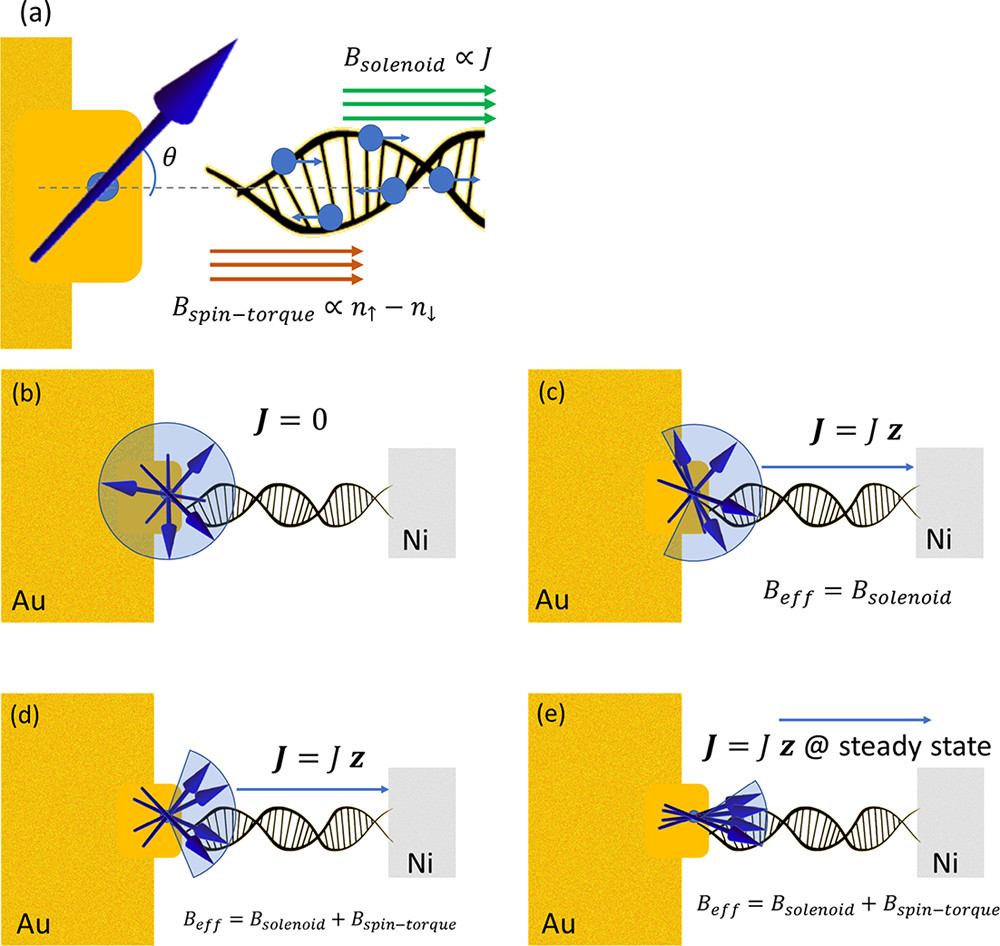}
    \caption{Alwan \& Dubi's (2021) dynamical on-bias interfacial model, depicting how a magnetic moment arises at substrate surface-states, bolstering spin-selectivity in CISS-junction setups. When bias is applied and charge-current is finite across the junction, a magnetic order is generated within the CISS molecule such that it exerts a solenoid-like magnetic field on the substrate surface. Given the magnetic polarizability of the substrate, this tilts magnetic dipoles of the surface states towards the molecular axis. This in turn further magnetizes the chiral bridge, and the resulting feedback mechanism is analogous to spin-transfer torque. Reprinted with permission from Alwan \& Dubi \cite{alwan_spinterface_2021}, Journal of the American Chemical Society \textbf{143}, 14235-14241 (2021). Copyright 2021 American Chemical Society.}
    \label{fig:fig16}
\end{figure}
Several coarse-grained models already attempt to account for CISS across both the transient and steady-state regimes \cite{fransson_charge_2022}, but they stop short of bridging the ‘strength gap’ in CISS. Here, Alwan \& Dubi\cite{alwan_spinterface_2021} propose that CISS’s anomalous spin-selectivity as experimentally observed arises from a spin-torque feedback mechanism between the substrate and the molecular bridge, using self-consistent mean-field calculations to calculate steady-state magnetocurrent anisotropy (Fig.~\ref{fig:fig16}). When an initial spin-current passes through the helical bridge, it sets up a small magnetic moment in the bridge, acting as an effective solenoid field that interacts with the initially randomly-aligned substrate surface orbitals. This aligns the orbitals, setting up an effective interfacial magnetic moment that increases the spin current, thus amplifying the effective interfacial magnetic field to strengths that can account for experimental levels of spin-selectivity in molecular junctions.

Within its purview, the spin-torque feedback mechanism can be employed to rationalize several experimental observations of CISS-transport \cite{dubi_spinterface_2022, alwan_temperature-dependence_2023, alwan_role_2024}. It of course does not provide any concrete explanation as to the origin of the effect, since CISS has been observed in substrates with negligible SOC \cite{adhikari_interplay_2023}. Observations of anomalous spin-selectivity upon incidence of linearly-polarized light in CISS-photoemission eludes a junction-based description, suggesting the chiral layer induces a persistent magnetic order in its substrates even at zero-bias. The latter was corroborated by observations of finite CISS-dependent Hall-response in a metal-substrate-SAM device without bias nor external magnetic field application \cite{ben_dor_magnetization_2017}, alongside direct observations of magnetization re-ordering and flipping upon adsorption of chiral molecules \cite{sukenik_correlation_2020, sharma_control_2020}. Remarkably, MIPAC experiments by Meirzada \textit{et al.} \cite{meirzada_long-time-scale_2021} then demonstrated that the induced magnetic-order is transient but can be long-lived, persisting over a period of days (the strength of which is correlated with the contact angle between chiral molecule and substrate).
\begin{figure}
    \centering
    \includegraphics[width=0.8\linewidth]{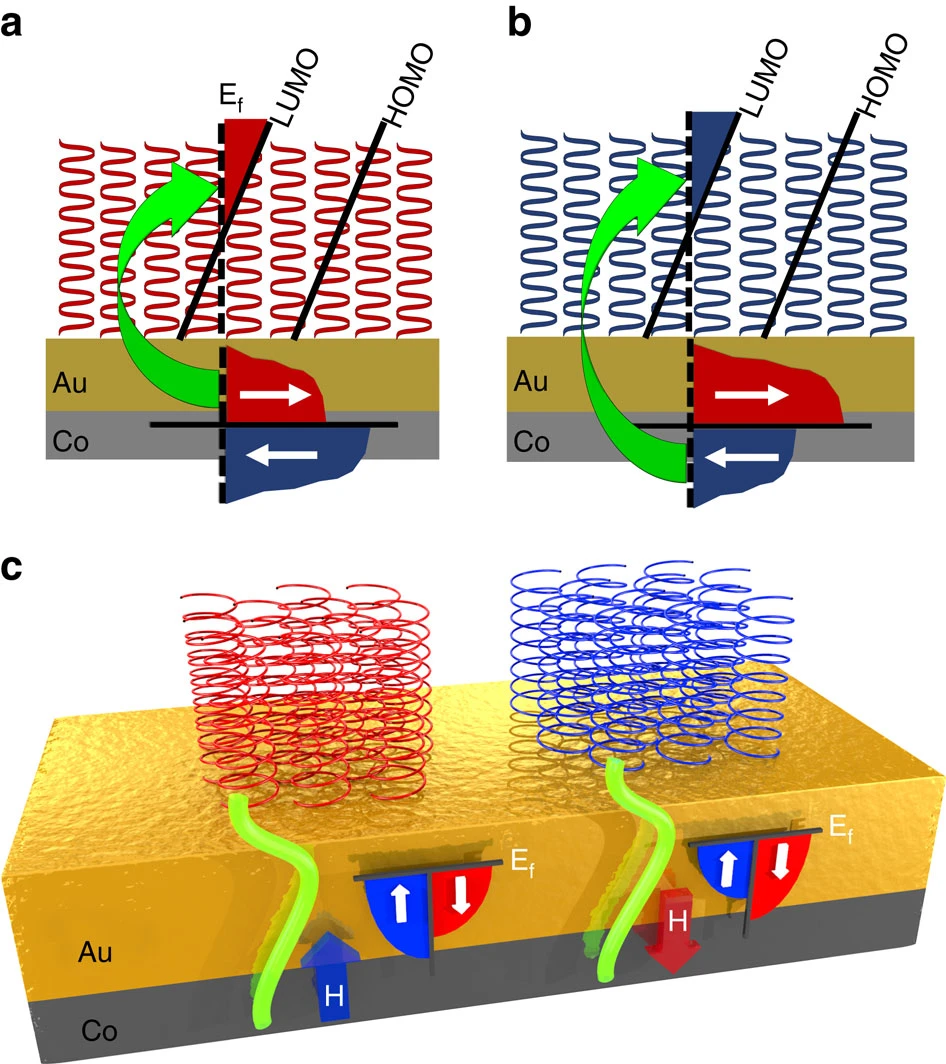}
    \caption{Ben Dor \textit{et al.} (2017) on-attach interfacial MIPAC model, describing how a magnetic order may be established on a substrate upon adsorption of a chiral film. Beginning from the premise that the chiral monolayer possesses a nontrivial dipole moment, charge transfer occurs to equalize the electrochemical potential of the adsorbed layer and the substrate surface, and this is spin-selective due to CISS in their account. One may relax this premise by instead accounting for interfacial wavefunction spread as offered by Zhao \textit{et al.} (2025). Reprinted with permission from Ben Dor \textit{et al.} \cite{ben_dor_magnetization_2017}, Nature Communications \textbf{8}, 14567 (2017). Copyright 2017 Authors, licensed under a Creative Commons Attribution (CC BY) license.}
    \label{fig:fig17}
\end{figure}
While analytic treatments of CISS’s stable non-equilibrium effects remain to be developed, heuristics offer possible explanations. For instance, Ben Dor \textit{et al.}\cite{ben_dor_magnetization_2017} propose that it owes to a spin-dependent proximity effect: the electron wavefunctions of the underlying ferromagnet delocalize into the SAM through the Au substrate (Fig.~\ref{fig:fig17}). The extent of penetration is spin-dependent due to the chiral layer, inducing an imbalance of spin-populations in the ferromagnet, thus setting a magnetic order; which is analogous to the charge-trapping model by Zhao \textit{et al.} previously detailed in Sec. \ref{part2C}. They make an equivalent semiclassical argument: the SAM dipoles lower the LUMO below the device’s Fermi level, inducing electrons to transfer from the FM, with stimulated spin-splitting due to CISS occurring in the process. Meirzada \textit{et al.} \cite{meirzada_long-time-scale_2021} elaborate on this, suggesting that large spin-exchange energies stabilize the transient magnetization \cite{dianat_role_2020}, with the order eventually decaying via intrinsic Gilbert damping and thermalization of spin states\cite{hickey_gilbert_2009, barron_false_2020}.

\begin{figure}
    \centering
    \includegraphics[width=0.8\linewidth]{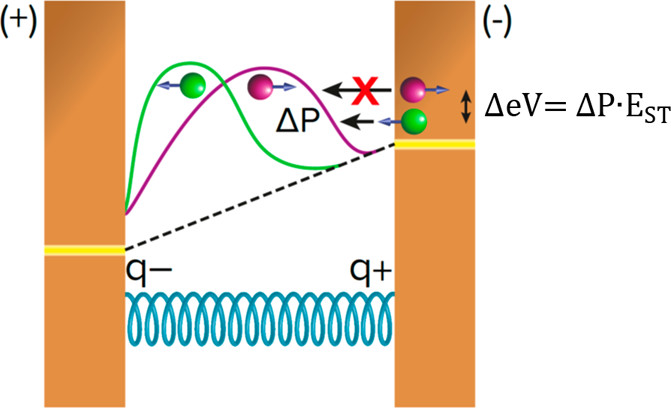}
    \caption{Illustration of Naaman \textit{et al.}'s (2020) spin blockade model. Spinor wavefunctions penetrate differentially through a dielectric chiral bridge, with the unfavored spin being localized closer to the donor, resulting in a spin-blockade forming for the unfavored spin (due to Pauli's exclusion principle). Reprinted with permission from Naaman \textit{et al.} \cite{naaman_chiral_2020}, The Journal of Physical Chemistry Letters \textbf{11}, 3660-3666 (2020). Copyright 2020 American Chemical Society, licensed under a Creative Commons Attribution (CC BY) license.}
    \label{fig:fig18}
\end{figure}

At this juncture, it is evident that charge-redistribution plays a crucial role in CISS-transport, one that is difficult to capture by general equilibrium models. Proposing an explanation for this, Naaman \textit{et al.} \cite{naaman_chiral_2020} offer the local spin-blockade picture, illustrated in Fig.~\ref{fig:fig18}.
The picture occurs in two stages: in-bridge spin-dependent charge-reorganization (SDCR) and into-bridge injection. Firstly, upon application of the bias field across a CISS-junction device, an electric dipole is set up within the chiral molecule, and charge-redistribution occurs. Due to CISS, the charge-redistribution is spin-dependent, such that electrons of the preferred spin delocalize further along the molecule than the non-preferred spin. Thus, the bridge sites closest to the donor electrode are preferentially populated by the non-preferred spin upon reorganization, giving rise to an interfacial “spin blockade”. In the next stage, an electron from the donor is injected into molecule, experiencing a spin-dependent tunneling barrier, the energy difference of which can be expressed as $\mathrm{SP}(V) \cdot \Delta E_{S,T}$. In words, the injection step is likened to the formation of either a singlet pair or a triplet pair, by sheer proximity, with the unpreferred ‘local’ spin at the interface. Preferred ‘excess’ spins from the donor form a singlet, while the unpreferred counterpart forms a triplet, with the energy difference being $\Delta E_{S,T}$. Thus, injected electrons not only encounter the initial spin-dependent barrier owing to the chiral bridge’s intrinsic SOC, it also faces an additional ‘spin blockade’ due to the singlet-triplet formation at the interface, drastically increasing spin-selectivity. Since the charge-reorganization step does not necessarily give rise to a ‘perfect spin-blockade’, a bias-dependent initial spin-separation index, $\mathrm{SP}(V)$, is used as a scale factor. 

In alignment with Zhao \textit{et al.} \cite{zhao25}, the spin-blockade picture accounts for observations of nonlinear response via the bias-dependent tunneling barrier, while magnetocurrent anisotropy readings are also predicted to fall at high bias voltages since the spin-blockade is no longer energetically comparable. It remains unresolved how the differences between the spin-blockade picture and the charge-trapping picture mean translate to their use in physical modeling. A finer outlook would require explicitly accounting for exchange interactions \cite{fransson_charge_2022}, highly-dielectric behavior, and possible long-range hopping or through-space contributions to transport. 

\subsection{CISS beyond Substrates: Chiral Transmission and Proximal Dispersion Forces}
\label{part3F}

One topic has remained conspicuously missing from our coverage thus far: how does CISS feature in biological processes? This topic has long been a central pillar in CISS focuses due to the as-yet unrealized potential of spin-manipulation in biochemical processes; but due in part to difficulty of experimental realization of CISS in biology \cite{qb25, gerhards2025rf}, insight into its effect on bare molecules is often derived indirectly. As such, much of our preceding discussion has centered around chiral molecules adsorbed on substrates. For all its promise, work on CISS-beyond-substrates is still nascent; but it has nevertheless already given plentiful insight on CISS, one part of which is its role in enantioselection \cite{fay2025enantioselectiveradicalreactionsinduced}.

\begin{figure}
    \centering
    \includegraphics[width=0.9\linewidth]{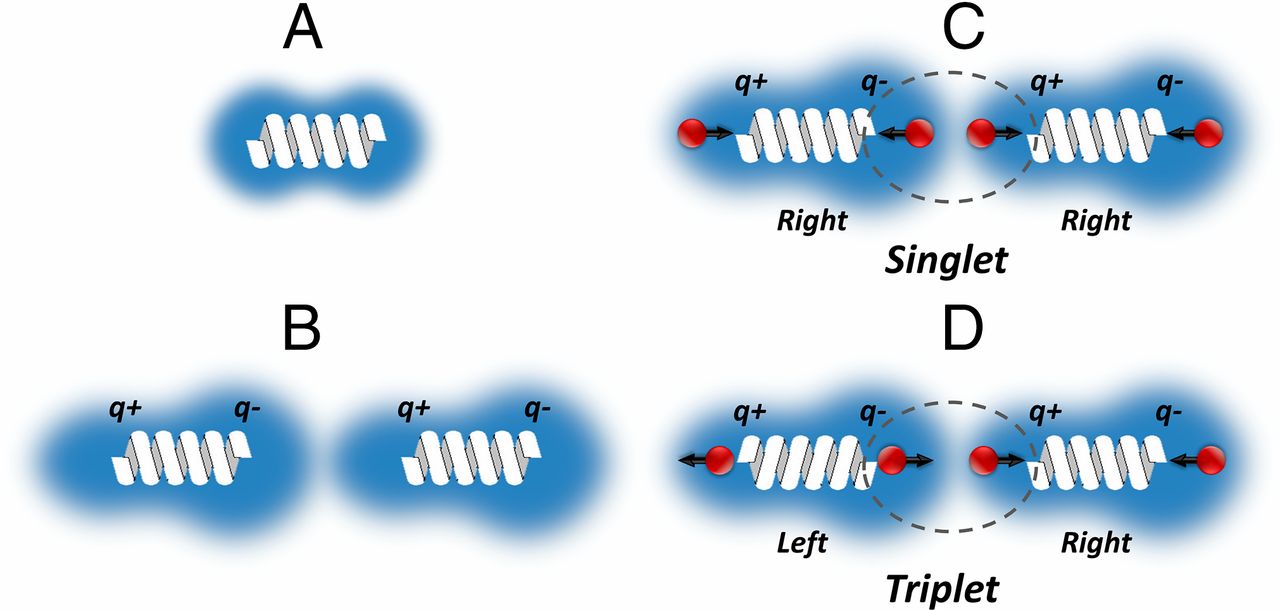}
    \caption{Illustration of Kumar \textit{et al.}'s capacitative spin-dependent charge-reorganization. Due to dispersion forces, induced dipoles are generated in two proximal chiral molecules, generating singlet or triplet states depending on the helical-sense matching. Reprinted with permission from Kumar \textit{et al.}\cite{kumar_chirality-induced_2017}, Proceedings of the National Academy of Sciences \textbf{114}, 2474-2478 (2017). Copyright 2017 National Academy of Sciences, licensed under a Creative Commons Attribution (CC BY) license.}
    \label{fig:fig19}
\end{figure}

For one, it is well-known that chiral macromolecular structures can spontaneously self-assemble from chiral monomers or achiral monomers in chiral media (see Liu \textit{et al.}\cite{liu_supramolecular_2015} for reference). Does CISS mediate this? Or specifically, can CISS occur between capacitatively-coupled molecules? Introduced by Michaeli \textit{et al.}\cite{michaeli_electrons_2016} and Kumar \textit{et al.}\cite{kumar_chirality-induced_2017}, capacitative SDCR was proposed to occur when two chiral molecules are brought into proximity with significant overlap of electronic wavefunctions (Fig.~\ref{fig:fig19}). In brief, the molecules interact via attractive dispersion forces, setting up an induced-dipole in each molecule and inciting charge-redistribution within each molecule. Due to CISS, the charge current is accompanied by spin-separation, forming singlet/triplet regions between the molecules if they are of the same/different chiral sense. This constitutes an enantioselective interaction that preferentially stabilize biomolecular-bonding due to different spin-correlation energies. However, calculations from Geyer \textit{et al.} \cite{geyer_contribution_2022}, peg the magnetic response from these fluctuations to be weaker than London forces, and Hedegard\cite{hedegard_chiral-induced_2022} elaborates how the effect is too weak to influence chemical reactions for SOC strength and dipole-matrix elements to order $\sim 10^{-3}$ eV, and excitation energies to order $\sim 1$ eV.

\begin{figure}
    \centering
    \includegraphics[width=\linewidth]{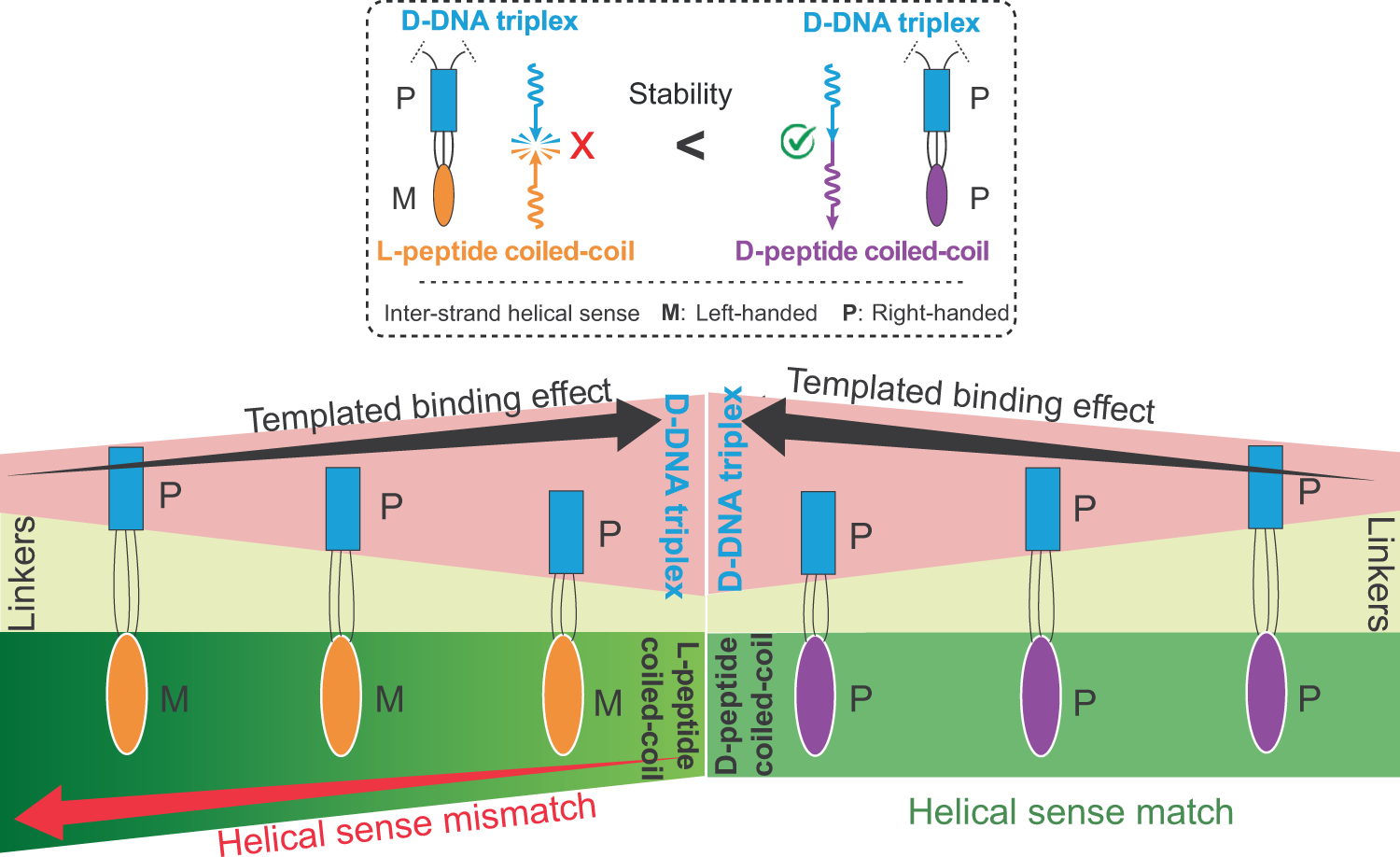}
    \caption{Graphical summary from Pandey \textit{et al.} on chiral-to-chiral communication propagating between two bio-macromolecular domains (DNA triple helices and peptides). The strength of the cross-strand linkage weakens with increasing linker length more drastically when the domains have different screw-sense helicity, suggesting a long-range electronic mechanism at play. Reprinted with permission from Pandey \textit{et al.} \cite{pandey_chirality_2022}, Nature Communications \textbf{13}, 76 (2022). Copyright 2022 Authors, licensed under a Creative Commons Attribution (CC BY) license.}
    \label{fig:fig20}
\end{figure}

While spin fluctuation-induced proximity effects may be negligible in chiral transmission, this does not preclude the presence of long-range interactions. For this, we may consider an alternative scenario: does CISS mediate bond-breaking of two covalently-bound chiral/helical molecules?

Remarkably, Pandey \textit{et al.} \cite{pandey_chirality_2022} demonstrated that helical-sense matching affects bonding strength even when the helical biomolecules are bonded through a linker and separated beyond the effective range of dispersion forces (Fig.~\ref{fig:fig20}). The mechanism behind this behavior is unclear, but we may indirectly take our cue from analyses focusing on with-substrate force spectroscopy: Kapon \textit{et al.}\cite{kapon_probing_2023}, studying atomic force-microscopy (AFM) CISS experiments, proposes that dispersion forces are supplemented by spin-exchange interactions, generated by the onset of polarized singlet states, giving rise to long-range magnetic interactions that correspond to chiral-discrimination in AFM settings. This result is further buttressed by Safari \textit{et al.} \cite{safari_spin-selective_2024}, in which significant magnetoconductance anisotropy was found for single-molecule AFM setups at low temperatures of 5K, precluding contribution from electron-phonon interactions and ensemble effects that appear as confounders for setups utilizing chiral nanolayers.

In light of these two perspectives, one may recognize proximal dispersion forces and exchange interactions are significant in chemisorption and enantioselectivity, but observations of these at work are facilitated by some form of contact, direct or indirect. Is this indicative of biomolecular CISS processes likely requiring indirect contact, or is there a possibility for appreciable non-local CISS effects in biology? And in either case, how can we better account for chiral transmission via self-assembly and broader biomolecular processes? 

\subsection{CISS in Spin Chemistry and Electron Transfer Reactions}
\label{sec:electron_transfer}

CISS has been proposed to be of relevance to several electron transfer reactions in biological systems, such as DNA, photosystem I, and cryptochrome, as well as artificial systems, such as quantum dots. While the underlying theory of electron transfer is well established \cite{nitzan_chemical_2006, may_charge_2011}, its interplay with CISS is still a developing field. In this section we explore some of the emerging models. Chiesa \textit{et al.}\cite{Chiesa_CISS_2021} defined a set of output states in proposed experiments to interrogate potential CISS effects in electron transfer processes through a chiral bridge linking a donor and acceptor pair. In particular, they identify a polarized state as arising from a spin filtering process through a chiral bridge with associated density matrix 
\[
\rho_{0}^{(\mathrm{F})} =
\begin{pmatrix}
0 & 0 & 0 & 0 \\
0 & \frac{1+p}{2} & 0 & 0 \\
0 & 0 & \frac{1-p}{2} & 0 \\
0 & 0 & 0 & 0 \\
\end{pmatrix},
\]
taken in the standard basis two-spin basis of $\{ \ket{\uparrow \uparrow},  \ket{\downarrow \uparrow}, \ket{\uparrow\downarrow}, \ket{\downarrow\downarrow} \}$ and satisfying $-1 \leq p \leq 1$ and $p\neq 0$. The final polarization of the acceptor here is given by $p=-2\mathrm{Tr}[\rho_{0}^{(\mathrm{F})}S_{z A}]$. By designing magnetic resonance experiments \cite{erSSb, erSS}, they aimed to distinguish outputs in terms of this polarized state, a singlet state, and a non-Boltzmann but unpolarized state, to determine whether the electron spin is polarized after electron transfer \cite{pyurbeeva22, pyurbeeva21} through a chiral bridge.  In addition to spin filtering mechanisms, spin coherent mechanisms have also been proposed. Luo \& Hore\cite{luo_chiral-induced_2021} formulated a phenomenological model, to investigate CISS effects in cryptochrome magnetoreception, that assumes spin polarization during the formation and recombination of a singlet born radical pair. Formally they arrive at the initial state 
\begin{equation}
    \ket{\psi_0} = \cos \left(\frac{\chi}{2}\right) \ket{S}  +  \sin \left(\frac{\chi}{2}\right) \ket{T_0},
\end{equation}
where $\ket{S}$ denotes the spin singlet state, $\ket{T_{0}}$ the spin triplet state, $\chi$ parametrizes the extent of spin-selectivity, and a recombination projection operator $P_{\mathrm{R}}^{(\mathrm{P})} = \ket{\psi_\mathrm{R}} \bra{\psi_\mathrm{R}}$, where
\begin{equation}
    \ket{\psi_\mathrm{R}} = \cos \left(\frac{\chi}{2}\right) \ket{S}  -  \sin \left(\frac{\chi}{2}\right) \ket{T_0}.
\end{equation}
In comparison to the model introduced by Chiesa \textit{et al.} the  corresponding density matrix is given by 
\[
\rho_{0}^{(\mathrm{P})} =
\begin{pmatrix}
0 & 0 & 0 & 0 \\
0 & \frac{1+\sin\chi}{2} & -\frac{\cos\chi}{2} & 0 \\
0 & -\frac{\cos \chi}{2} & \frac{1-\sin \chi}{2} & 0 \\
0 & 0 & 0 & 0 \\
\end{pmatrix}.
\]
Using this formalism, Luo \& Hore demonstrated that external magnetic field sensitivity can be enhanced in radical pairs subject to strongly asymmetric recombination. Further studies by Tiwari, Raghuvanshi \& Poonia\cite{tiwari_quantum_2023, tiwari_role_2022} extended upon this result and suggested that CISS may amplify the relative entropy of coherence for both the electronic subsystem alone and the combined electronic-and-nuclear system, and have applied the model to planaria regeneration\cite{Tiwari_planaria_2024}. Nonetheless, open questions remain for the role of CISS in electron transfer reactions in biophysical processes. These include identifying the correct theoretical framework and establishing if predictions are consistent with previous observations, and experimental verification of CISS effects, for which interpretations of EPR spectra of photosynthetic bacterial reaction centers revealed no significant CISS generated polarization\cite{Ren_EPR_2023}.  

\begin{figure}
    \centering
    \includegraphics[width=0.8\linewidth]{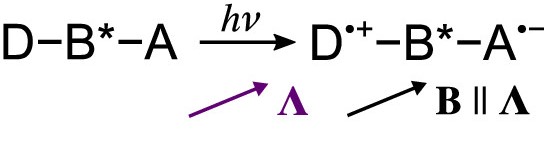}
    \caption{Simple schematic of the radical-pair mechanism: when a donor and acceptor molecule, covalently linked by some chiral bridge, is photoexcited to generate a radical pair, the system then forms a charge transfer state. Direct diabatic coupling provides singlet character, while SOC ($\Lambda$) contributed by the chiral bridge provides triplet character, with chirality altering the phase sign. No spin-polarization emerges, but instead spin-coherence. Application of a magnetic field parallel to the SOC breaks time-reversal symmetry of the Hamiltonian and allows for decay of coherence, which can then be experimentally probed by electron paramagnetic resonance setups. Adapted with permission from Fay\cite{fay_chirality-induced_2021}, The Journal of Physical Chemistry Letters \textbf{12}, 1407-1412 (2021). Copyright 2021 American Chemical Society.}
    \label{fig:fig9}
\end{figure}

Another spin coherent mechanism was introduced by Fay \cite{fay_chirality-induced_2021} who demonstrated that, based on a microscopic derivation for a one-step electron transfer in radical pairs, chirality in conjunction with SOC generates quantum coherence between singlet and triplet spin-states, yet does not generate spin polarization; we briefly detail the model as illustrated in Fig.~\ref{fig:fig9}.Consider a charge-transfer reaction for a photoexcited radical pair system. The excited precursor state can be designated $\ket{0} = \ket{DA^*}$, and the charge-separated radical-pair state can be designated $\ket{1} = \ket{D^{\bullet+}A^{\bullet-}}$, where $D$($A$) indicates the donor(acceptor). Of note, $\ket{0}$ only exists in the singlet configuration while $\ket{1}$ can exist in either singlet and triplet configuration. The Hamiltonian for such a system dressed in an SOC field is thus: 
\begin{equation}
    H = H_0 \Pi_0 + H_1 \Pi_1 + V,
\end{equation}
where $H_m \Pi_m$ denote the Hamiltonian and projection operators onto charge-transfer state $m$, with $H_m = H_{ms} + H_{mn}$ accounting for both electron and nuclear spin subspaces while $\Pi_0 = \ketbra{0}{0} \hat{P}_S$ and $\Pi_1 = \ketbra{1}{1} (\hat{P}_S + \hat{P}_T)$ account for each state's allowed spin-multiplicity through the singlet(triplet) projection operators $\hat{P}_S$($\hat{P}_T$). $V$ consists of our potentials of interest, the spin-conserving diabatic coupling $V_{DC}$ and SOC:
\begin{equation}
    \begin{split}
        V &= V_{DC} + V_{SOC} \\
        &= \Delta \left( \hat{P}_S\ketbra{0}{1}+ \ketbra{1}{0} \hat{P}_S \right) - i \left( \hat{P}_S\ketbra{0}{1} \pmb{\Lambda} \cdot \hat{\mathbf{S}}_1 - \pmb{\Lambda} \cdot \hat{\mathbf{S}}_1 \ketbra{1}{0}\hat{P}_S \right) \\
        &= \Gamma \left( \hat{P}_S\ketbra{0}{1} \hat{U}^{\dagger} + \hat{U} \ketbra{1}{0}\hat{P}_S \right),
    \end{split}
\end{equation}
with $\pmb{\Lambda}$ being a real-valued momentum-dependent vector coupling and $\hat{\mathbf{S}}_1$ being the spin-operator on a particular electron denoted `1' in the electron-transfer process, giving $\hat{U}=(\Delta + i \pmb{\Lambda} \cdot \hat{\mathbf{S}}_1)/\Gamma$ and $\Gamma = \sqrt{\Delta^2 + |{\pmb{\Lambda}}|^2 /4}$.

Physically, $\hat{U}$ acts to rotate electron spin about an axis $\hat{\mathbf{n}} = \pmb{\Lambda}/|{\pmb{\Lambda}}|$ for an angle $2\theta = 2\cdot \arctan \left( |\pmb{\Lambda}|/2\Delta\right)$. Additionally, opposite enantiomers, denoted by $+$ and $-$, have opposite $\mathrm{sgn}(\pmb{\Lambda})$ such that $\pmb{\Lambda}_+ = -\pmb{\Lambda}_-$. Proceeding to solve the Nakajima-Zwanzig equations for the system, under the Markovian assumption and that the reaction is instantaneous and irreversible (i.e. $k_b \rightarrow 0$), Fay obtains:
\begin{equation}
    \ket{\psi_0^{\pm}} = \cos \theta_+ \ket{S} \pm i \sin \theta_+ \ket{T_0(\mathbf{n}_+)},
\end{equation}
$\ket{T_0 (\mathbf{n}_+)}$ the triplet-state defined along the axis $\mathbf{n}_+$, and $\ket{\psi_0} = \hat{U}\ket{S}$ the initial spin state of the system. In comparison to the previous initial states Fays model gives the initial density matrix
\[
\rho_{0}^{(\mathrm{C})} =
\begin{pmatrix}
0 & 0 & 0 & 0 \\
0 & \frac{1}{2} & -\frac{1}{2}(\cos\chi + \mathrm{i} \sin\chi) & 0 \\
0 & -\frac{1}{2}(\cos\chi - \mathrm{i} \sin\chi)  & \frac{1}{2} & 0 \\
0 & 0 & 0 & 0 \\
\end{pmatrix}.
\]

\begin{figure}
    \centering
    \includegraphics[width=0.8\linewidth]{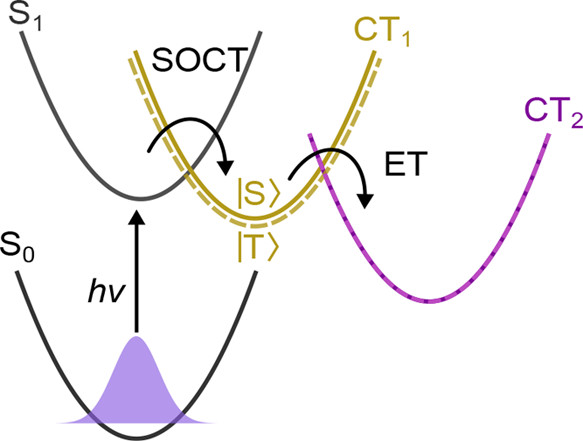}
    \caption{Schematic for conversion of chiral-induced spin-coherence to CISS via a final incoherent step. In realistic conditions, the radical-pair undergoes multiple downhill electron transfers, resulting in a final well-separated ion pair. This can be modeled by adding a final dissipative or incoherent electron transport step that generates spin-selection from spin-coherence. Reprinted with permission from Fay \& Limmer \cite{fay_origin_2021}, Nano Letters \textbf{21}, 6696-6702 (2021). Copyright 2021 American Chemistry Society.}
    \label{fig:fig10}
\end{figure}

Effectively, in the absence of any incoherent (i.e. dissipative) process, Fay’s radical pair model demonstrates that CISS is only spin-selective for coherence and not for polarization, a conclusion in line with analysis in Sec. \ref{part2A}. Fay \& Limmer \cite{fay_origin_2021} then extended upon this result by considering a two-step model to describe CISS-photoemission (Fig.~\ref{fig:fig10}), likening the process to chemically-induced dynamic electron-polarization\cite{ayscough_chemically_1979}. In brief, by introducing an additional exchange coupling term between donor and acceptor spins in the radical pairs, the triplet state now has an oscillating real part; this oscillating spin polarization is subsequently converted into a static quantity via a final incoherent electron-transfer step. Furthermore, the model was shown to have a temperature dependence consistent with experimental results by Carmeli \textit{et al.} on CISS in electron transfer in photosystem I \cite{carmeli_spin_2014}. In a subsequent work, Fay \& Limmer explored the role of CISS in charge recombination, and a reaction operator was derived that described superexchange and incoherent hopping limits \cite{fay_donorBridgeAcceptor_2023}. Most recently, this framework has also been extended to demonstrate enantioselectivity via CISS and thereby give rise to new methods for asymmetric synthesis using spin-polarized electrons \cite{fay2025enantioselectiveradicalreactionsinduced}. 

Although, the CISS models that generate polarization or only coherence give rise to distinct initial states, by combining a local phase rotation operation on a transferred electron in combination with evolution under exchange coupling, Smith \textit{et al.}\cite{ciss25rp} have provided a unified scheme that allows interpolation between these models. Using this, they have also extended to accommodate triplet precursor radical pairs, allowing effects due to spin selectivity in polarization or coherence to be identified and the underlying conditions for enhancements in magnetic field sensitivity and coherence to be elucidated in radical pairs in cryptochrome. 
\begin{figure}
    \centering
    \includegraphics[width=\linewidth]{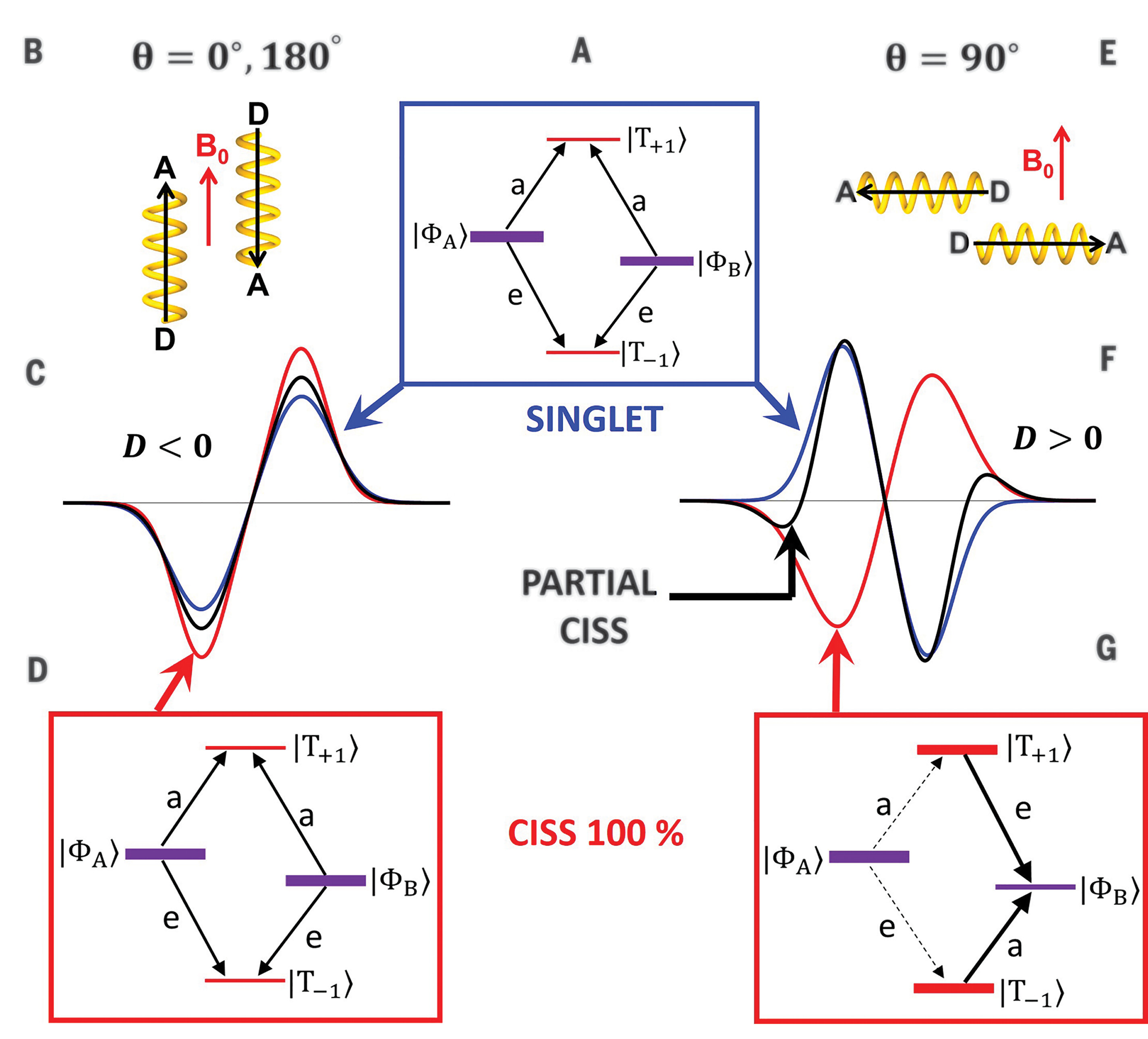}
    \caption{The effect of CISS on the spin states of spin-correlated radical pairs (SCRPs), as revealed by time-resolved electron paramagnetic resonance (TREPR). (A) In the absence of CISS, a SCRP formed in a singlet state evolves in the presence of a strong static magnetic field $\bf{B}_{0}$. (B, E) Alignment of the chiral molecules parallel ($\theta = 0^{\circ}, 180^{\circ}$)  or perpendicular ($\theta = 90^{\circ}$) to $\bf{B}_{0}$. (C, F) TREPR spectra showing singlet-born SCRP signals with and without CISS contributions, where red traces indicate full CISS generated spin polarization, black traces indicate partial CISS contributions, and blue the case without CISS. (D, G) Schematic diagrams of spin states under 100\% CISS contribution. $D$ represents spin-spin interactions. Reprinted with permission from Eckvahl \textit{et al.}\cite{eckvahl_direct_2023}, Science \textbf{382}, 197–201 (2023). Copyright 2023 Authors. }
    \label{fig:eckvahl}
\end{figure}
Recently, Eckvahl \textit{et al.} \cite{eckvahl_direct_2023} produced evidence of a direct observation of CISS in intramolecular electron transfer for an isolated, photoexcited chiral molecule within a liquid crystal medium. Specifically, they used time-resolved electron paramagnetic resonance (TREPR) spectroscopy to identify signatures of the CISS effect on the spin states of spin-correlated radical pairs by rotating the molecules relative to an applied magnetic field (Fig.~\ref{fig:eckvahl}). While the model of Luo \& Hore is used for comparison, the authors state that the models discussed in this section, i.e. associated with $\rho_{0}^{(\mathrm{F})}$, $\rho_{0}^{(\mathrm{P})}$, and $\rho_{0}^{(\mathrm{C})}$, are equivalent for describing the TREPR spectra provided that polarization effects are incorporated. However, by comparing these density matrices, we see that the models are not equivalent and differ with respect to coherence elements. Although not relevant for the interpretation of TREPR, these differences could potentially be of importance for applications utilizing quantum resources, such as quantum sensing or quantum information protocols like quantum teleportation \cite{wasielewski_spinChemQISReview_2023, chiesa2025ciss}. In Sec. \ref{sec:scrp_platforms}, we revisit model differences in CISS to discuss a scenario in magnetic field sensing where stark differences can arise if CISS is selective for polarization or coherence only. 

Beyond the donor-bridge-acceptor context, quantum effects or enhancements have also been discussed in the context of CISS-transport through junctions. Wang \textit{et al.} \cite{wang_spin_2021} identify that spin-polarization in CISS-transport can be induced and amplified by the presence of Fano resonances, arising from self-interference induced by a non-equilibrium distribution of injected states (i.e. optically excited) and further compounded by contributions from quasi-degenerate energy levels common in complex molecules \cite{exci14}. In a similar vein, collected efforts from the Subotnik group \cite{bian_modeling_2021, bian_meaning_2022, wu_electronic_2021,teh_spin_2022} raise the striking argument that failure of the nuclear-electronic adiabatic assumption can be a major contributor to CISS. In the vicinity of conical intersections, quasi-degenerate states can experience a significant Berry force, physically interpretable as a nuclear screening force acting on the electron\cite{chandran_electron_2022}, that can severely affect pathway selection of quantum states and potentially give anomalous spin-selectivity for small SOC strengths. Taken together, the results of this section highlight the substantial effects CISS may have on electron transfer reactions, and its observation at the molecular level demonstrates its potential in quantum technologies. 

\section{Bridging the Gap to Emerging Applications: Biomimetics, Quantum Information and Sensing}
\label{part4}
Given the widespread manifestations of the CISS effect and its potential for spin manipulation, it is increasingly recognized not only for its fundamental significance but also for its potential in emerging quantum and molecular technologies, especially considering its reputed robustness at ambient temperatures. Applications span spintronics, including spin filters, spin valves, and spin-selective tunneling junctions, as well as chiral-induced spin light-emitting diodes (spin-LEDs), molecular switches, and enantioselective catalysts. Beyond device applications, CISS has also attracted attention in the context of biological electron transfer and the prospect of coherent control of chemical reactions. In this section, we maintain a focused discussion on sensing applications, quantum information and control concepts, and insights from biology \cite{Lambert2013}, culminating in emerging spin-correlated radical pair platforms. Broader appraisals of CISS-enabled applications and implications are available in recent literature \cite{bloom_chiral_2024, firouzeh_compositeMaterialsDevices_2024, jiawei_review_chiralNanomaterials_2025, naaman_review_spinChem_2020, naaman_biological_2022}. 

\subsection{CISS-Inspired Sensing and Detection} 
Several proposed applications have emerged that harness the CISS effect’s ability to polarize electron spins, enabling novel sensing and detection schemes. One promising route involves the utilization of magnetoresistive readouts and magnetoresistance properties of chiral systems \cite{yang_linear_2021, kang_roomTempSpinValve_perovskite_2024, kondou_thermallyDriven_2022, Aragones_magnetoresistance_2022, huisman_inelastic_2021, tirion_hanle_2025} for spintronic sensor development \cite{byeonghwa_review_magnetosensors_2022}. A particularly active area involves bioelectronic platforms \cite{Bostick_2018}, where chiral molecules such as DNA hairpins \cite{latawiec25}, peptides, or oligopeptides, serve as recognition elements along with spin-sensitive detection. These platforms present promising opportunities for biomedical diagnostics, offering spintronic alternatives to traditional label-based methods. In these efforts, Wang and co-workers proposed the use of a self-assembled monolayer of polypeptides and a CISS constructed Kerr technique to modulate incident linear polarized light on a non-magnetic metal surface with chiral molecules, where the deflection angle depends on the chirality of the molecule \cite{zhang_kerr_2023}.
\begin{figure}
    \centering
    \includegraphics[width=.95\linewidth]{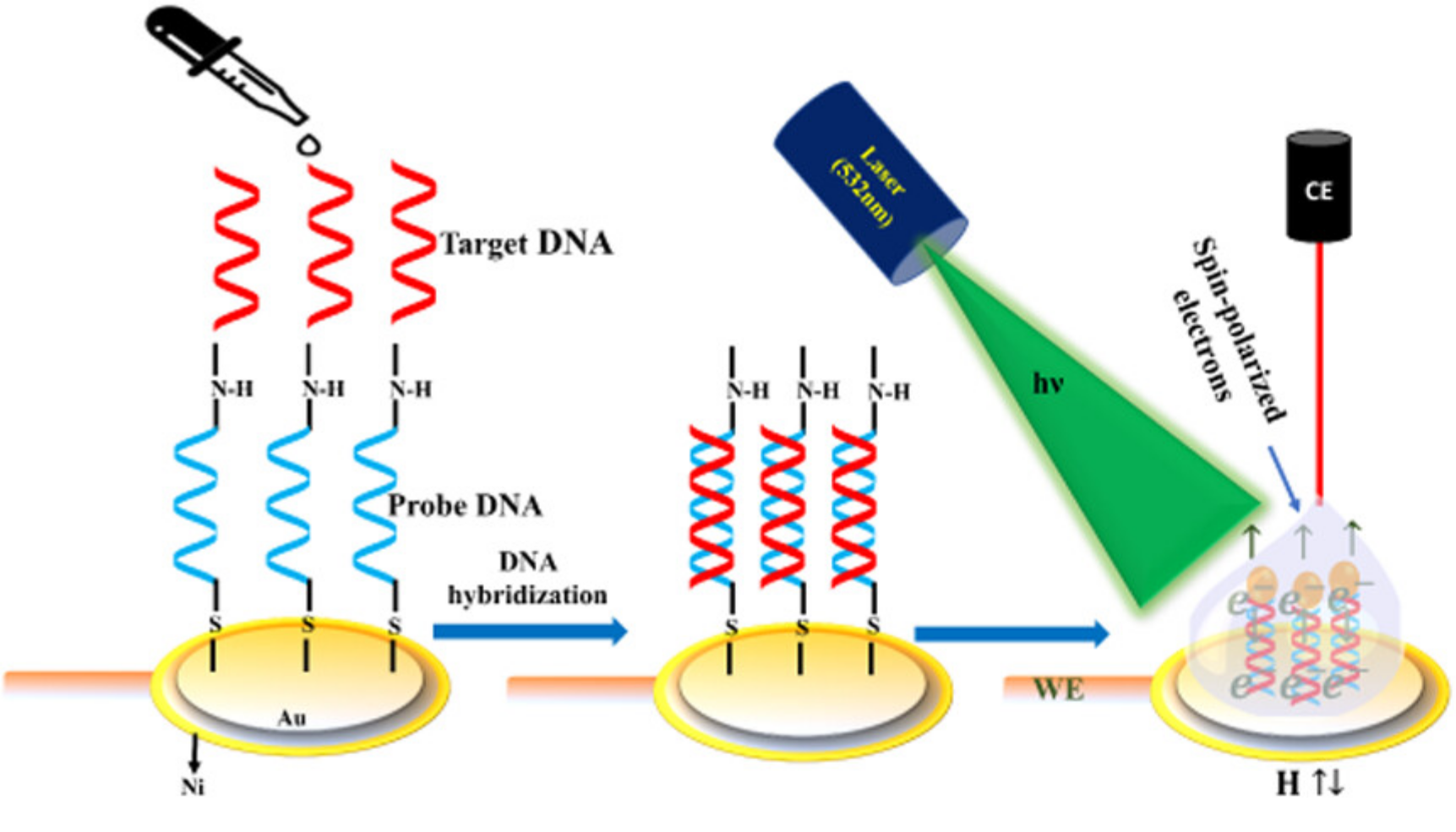}
    \caption{Schematic of a CISS-based DNA hybridization sensor employing a quantum dot–DNA device. A probe single-stranded DNA (ssDNA) is immobilized on a gold-coated Ni substrate, followed by hybridization with target DNA. Upon illumination with circularly polarized light ($\lambda = 532$ nm), spin-polarized electrons are generated and transported through the hybridized double-stranded DNA (dsDNA). The CISS effect induces a differential change in charge-transfer resistance ($\Delta$Rct) between spin-up and spin-down electrons for both ssDNA and dsDNA, enabling spin-dependent detection of hybridization. This spin-selective readout enhances sensitivity over conventional charge-based methods and enables a detection limit as low as 10 fM under illumination. Adapted with permission from Bhartiya \textit{et al.}\cite{bhartiya_dnaHybridisation_2023},  Anal. Chem. \textbf{95}, 3656–3665 (2023). Copyright 2023 American Chemical Society.}
    \label{fig:bhartiya}
\end{figure}

DNA-based schemes are also undergoing rapid development, including investigations into CISS effects in Quantum Dot-DNA systems \cite{bhartiya_dnaHybridisation_2023, bangruwa_sequenceControl_2023} (Fig.~\ref{fig:bhartiya}), and CISS-based DNA hybridization sensors \cite{bangruwa_dnaHybridisation_Detection_2024, bhartiya_dnaHybridisation_2023, tiwari_dnaHybridisation_2024}, which can be used to measure genetic similarity and finds several applications including drug discovery, pathogen detection, and hereditary disease diagnosis. Related strategies have demonstrated the detection of UVC-induced DNA damage via spin polarized electron signals \cite{bangruwa_dnaDamage_2022}, potentially offering greater robustness over conventional electrochemical biosensors. Closely related is the concept of CISS-enabled enantioselective detection, where spin-polarized electron transport is influenced by the chirality of a target molecule, enabling enantiomer differentiation. Recent studies have explored spin-dependent enantiomeric separation \cite{banerjeeGhosh2018, banerjeeGhosh2020} and the use of nuclear magnetic resonance (NMR) as a tool for chiral sensing \cite{Georgiou2024}, with implications for chemical analysis and quantum information science. 

\begin{figure} [h]
    \centering
    \includegraphics[width=\linewidth]{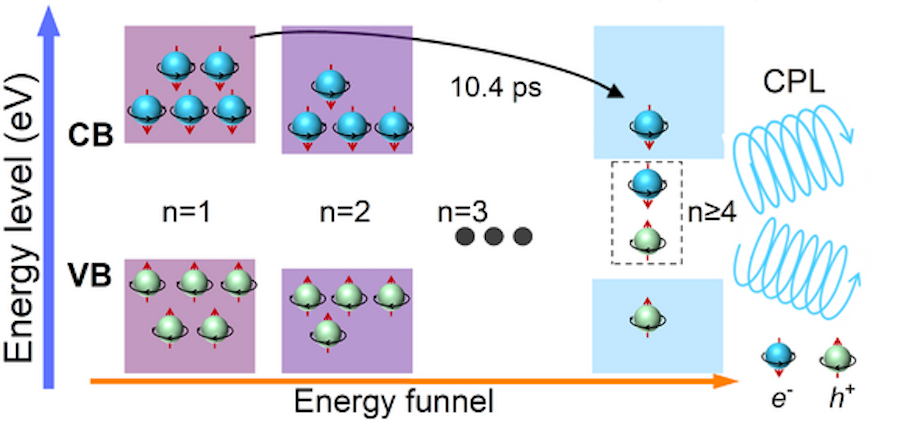}
    \caption{A pictorial illustration for spin polarized exciton transfer and recombination in a high‐performance sky‐blue perovskite spin‐LED due to chiral ionic liquid implantation and passivation. Adapted with permission from Zhang \textit{et al.}\cite{zhang_spinOpto}, Adv. Funct. Mater., 2503088 (2025). Copyright 2025 Wiley.}
    \label{fig:ciss_led}
\end{figure}

Spin-selective photoluminescence presents another route for CISS-based sensing. Chiral organic emitters and perovskites can exhibit circularly polarized luminescence (CPL) \cite{mishra2024, jingying_perovReview_2023}, traditionally used to investigate chirality. However, CPL may also serve as a spin-based optical output in future sensor architectures (Fig.~\ref{fig:ciss_led}). Paralleling this, chiral hybrid perovskite systems have also seen use as spin light-emitting diodes \cite{zhang_spinOpto, pan_spinOpto, He2025, tang_spinOpto, yang_spinOpto}, with Tang \textit{et al.} reporting the first single-junction perovskite spin-LED operating at room temperature \cite{tang_spinOpto}, and with chiral SOC field strength later being shown to explicitly correlate with both external quantum efficiency and degree of circular polarization in electroluminescence, revealing a magneto-chiroptical interplay that can be tuned by compositional engineering \cite{yang_spinOpto}. Overall, a number of studies have quantified the strength of chirality-induced spin-orbit coupling (CISOC) and its correlation to dimensionality, local inversion asymmetry, and electronic delocalization across chiral perovskites, thus paving the way for harnessing the CISS effect towards the development of practical spin–optoelectronic device architectures \cite{yang_spinOpto, Kattnig2009}. 

A theoretical basis for spin‑polarized injection and circularly polarized electroluminescence in chiral perovskite spin‑LEDs could be provided by modeling a spinor confined to a helical path, as considered recently by Ventra \textit{et al.}\cite{diventra2025}, that naturally acquires a geometric chirality‑induced spin–orbit coupling, expressed as $H_\chi = \alpha_\chi\,\hat{\boldsymbol\sigma}\cdot\hat{\mathbf p}$, with \(\alpha_\chi\propto \kappa/R\) (over some curvature \( \kappa \) and radius \( R \)), enabling spin–momentum locking and breaking symmetry between \(T^\uparrow\) and \(T^\downarrow\) (which are the transmission probabilities for spin-up and spin-down electrons, respectively) even in light‑atom systems. This is further bolstered by findings from Zhang \textit{et al.} \cite{guosun_2025} showing that asymmetric spin‐velocities lead to a nonzero steady‐state polarization \(P_{\rm stat}=(n_\uparrow-n_\downarrow)/(n_\uparrow+n_\downarrow)\), even without external magnetic fields.
In fact, from a quantum thermodynamic perspective, this implies that the state transformation mechanisms underlying the CISS effect could be acting as passive isothermal quantum erasers or initializers that are robust at room-temperature. In treating CISS as a quantum channel encoding spin-selective transmission amplitudes, chiral asymmetry due to molecular geometry, and open quantum systems effects like dephasing \cite{guosun_2025, dephasing23}, one can conceptualise formally recasting it as an energetically optimal qubit transformer \cite{qTr25, dunlop24}, thereby putting it in the same family as Landauer erasers, Maxwell’s demon type quantum engines \cite{phkv-wrsd}, and information-powered refrigerators.

Considered together, these developments demonstrate the versatility of the CISS effect for spin-selective sensing. Given the rapid growth and depth of the field, we have elected to highlight a few of these emerging sensing platforms, rather than an exhaustive coverage, which we hope provides some insight into the breadth of the field and potential applications, such as biomedicine \cite{barron2025, Wu_biomedical_2025, aiello24weak}. Further insight into quantum information and technology device implementation prospects of CISS can be found in several reviews \cite{firouzeh_compositeMaterialsDevices_2024, jiawei_review_chiralNanomaterials_2025, chiesa_review_quantum_2023, aiello2022, wasielewski_spinChemQISReview_2023, mishra_spintronicsReview_2025}. While many of these systems have not yet been investigated for active utilization of quantum resources, such as entanglement or coherence, their underlying quantum structure offers potential for quantum-enhanced sensing platforms \cite{Wu_biomedical_2025}. In the following section we examine an emerging platform based on spin-correlated radical pairs, and explore the relation between quantum resources and sensing applications. 

\subsection{Spin-Correlated Radical Pair Platforms} 
\label{sec:scrp_platforms}
Previously the potential for spin-correlated radical pairs (SCRPs) to be used as molecular qubits for quantum sensing \cite{xie_SCRPsensing_2023} has been recognized and covered in a comprehensive review \cite{mani_review_2022}. Here we will summarize some of the key points, elaborate on connections with CISS in radical pair systems, and finalize with a discussion on prospects for SCRPs generated in donor-chiral bridge-acceptor molecules as an emerging platform for quantum information and sensing.

In the previous section we described several sensing platforms that make use of the CISS effect. Following the classifications of Degen \textit{et al.}\cite{Degen_sensing_2017}, most of these predominantly fall under the category of type-I sensors, which make use of a quantum system to measure a classical or quantum physical quantity. However, there is potential scope for type-II sensors, that make use of quantum coherence to measure a quantity, or type-III sensors that make use of entanglement to improve the sensitivity or precision of measurement beyond what is classically possible. 
\begin{figure}
    \centering
    \includegraphics[width=\linewidth]{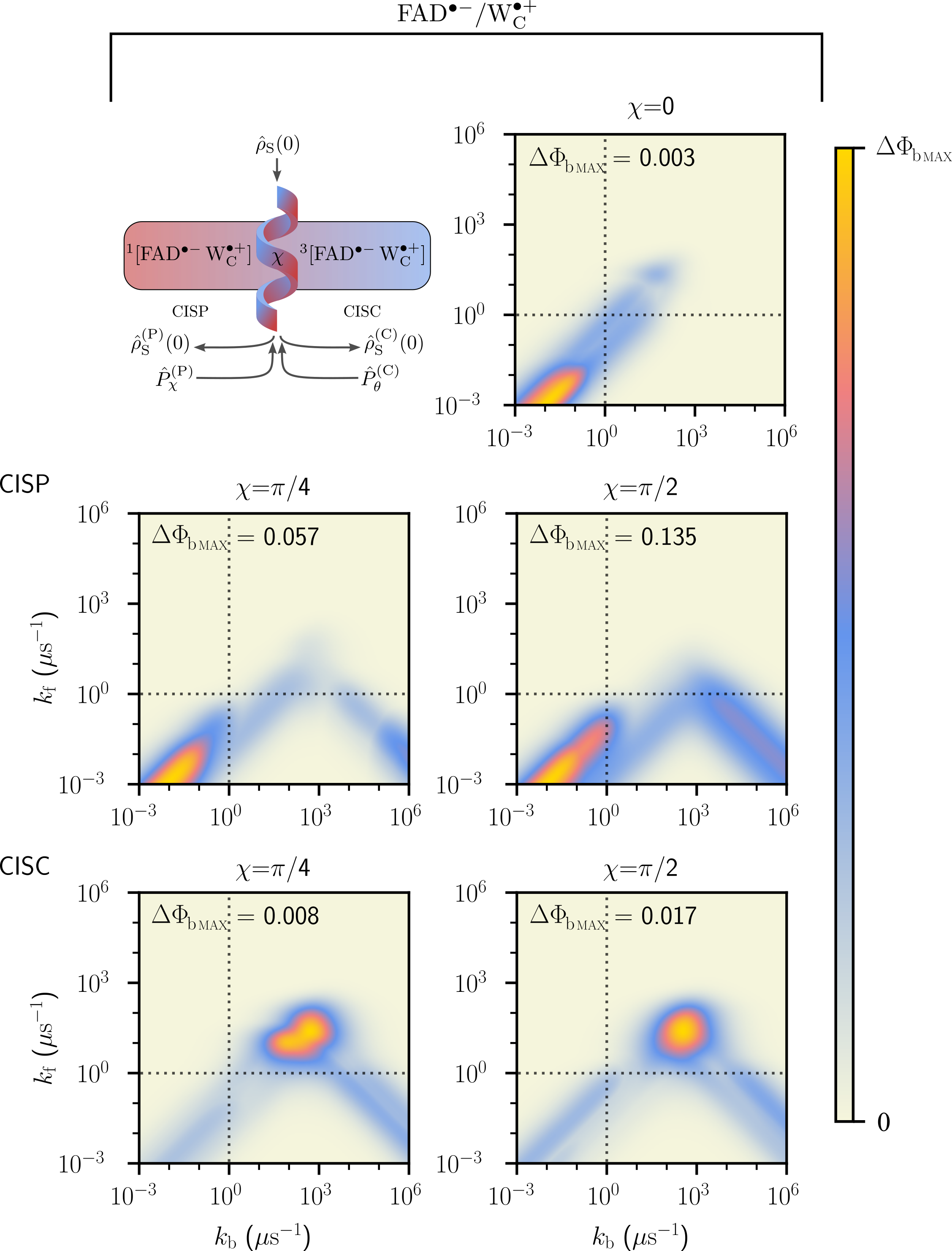}
    \caption{CISS modulated magnetic field sensitivity for the flavin-tryptophan (FAD$^{\bullet -}$/W$_{\mathrm{C}}^{\bullet+}$) radical pair system. Heatmaps show the sensitivity as assessed by the difference in recombination yields ($\Delta \Phi_\mathrm{b}$) sampled over $300$ magnetic field orientations and $200\times 200$ combinations of recombination $k_{\mathrm{b}}$ and forward reaction $k_{\mathrm{f}}$ rates, under Zeeman, hyperfine, and electron-electron dipolar interactions. The case of no CISS effects $\chi=0$, is compared to varying degrees of CISS $\chi=\pi/4$ and $\chi=\pi/2$ for both chirality induced spin polarization (CISP) and chirality induced spin coherence (CISC) models. Significant sensitivities are produced for asymmetric recombination, indicative of the quantum Zeno effect, only if polarization is generated. Dotted lines denote conventional parameter choices of reaction rates at $1\,\mu$s$^{-1}$. Adapted with permission from Smith \textit{et al.}\cite{ciss25rp}, AVS Quantum Science \textbf{7}, 032601 (2025). Copyright 2025 Authors, licensed under a Creative Commons Attribution (CC BY) license.}
    \label{fig:ciss_zeno}
\end{figure}
In particular, SCRPs have a few desirable properties in which coherent singlet-triplet interconversion can last up to $100$ ns and potentially up to microsecond timescales in radical pairs relevant to magnetoreception \cite{rporis25, chowdhury2024probing, hiscock19, Nelson_hyperfine_2018, nielsen17}. This makes them promising candidates for quantum sensing applications and beyond. For example, Xie \textit{et al.}\cite{xie_SCRPsensing_2023} used a photogenerated SCRP formed from a radical ion pair to sense an electric field change created at one radical ion of the pair using molecular recognition. Using pulse-EPR measurements of oscillations caused by the distance dependent electron-electron dipolar coupling to analyze the distance between the two spins of the SCRP. Although SCRPs have currently been limited to type-I and type-II sensors, given that they constitute molecules that have unpaired electron spins that can be generated in pure quantum states, there is a potential to make use of quantum properties such as entanglement and efforts underway to make use of them as spin qubit pairs. Beyond this, Lin \& Mani \cite{lin_tripletAnnihilation_2025} have explored their use in amplifying magnetic field effects via triplet-triplet annihilation and the Wasielewski group has made progress into the application of SCRPs in quantum information and communication protocols \cite{poh25, Brown_review_2024, wasielewski_spinChemQISReview_2023, harvey_SCRP_2021, Rugg2019}, and has reviewed the trajectory to these applications from photosynthetic energy transduction and biomimectics. 

On the topic of attaining inspiration from biochemical design principles, and optimizing these in new technologies, SCRPs have been studied heavily in the context of the spin biochemistry of magnetoreception. Quantum phenomena, such as quantum coherence and superposition, are generally not expected to survive in noisy ambient conditions, and so it is of interest to identify if nature has found ways to protect quantum resources in biological settings \cite{aplQ}. Fay \textit{et al.} have identified the radical pair requires full quantum mechanical calculations for an accurate description, suggesting that quantum effects may play an important role \cite{fay_how_2020}. Moreover, investigations into radical pair-based magnetosensitivity may provide useful insights and tools for SCRP applications, such as elucidating the utilitarian role of quantum resources \cite{Smith_obs_2022, Rakshit_sustained_2021, Kominis_rel_ent_2020, Cai_chemical_compass_2013, Gauger_sustained_2011}, quantum limit frameworks to assess quantum sensing capabilities\cite{Kominis_energy_resolution_2025, opt24, Guo_qfi_2017, Vitalis_quantum_limited_2017}, quantum control of SCRPs\cite{Tateno_RYDMR_2025, hore2025spin, prxQC, Cai_control_2010}, and enhancement mechanisms for magnetic field sensitivity and coherence, e.g. through spin relaxation \cite{Jiate_decoherence_2024, Kattnig_relax_enhance_2016}, radical motion\cite{Smith_driven_2022, contQFIdr25}, three-radical mechanisms\cite{Babcock_scavenger_2021,Keens_dipolarly_2018, Kattnig_scavenger_2017}, and the quantum Zeno effect \cite{ciss25rp, Dellis_zeno_2012, kominis_quantum_2009}.

\begin{figure}
    \centering
    \includegraphics[width=.75\linewidth]{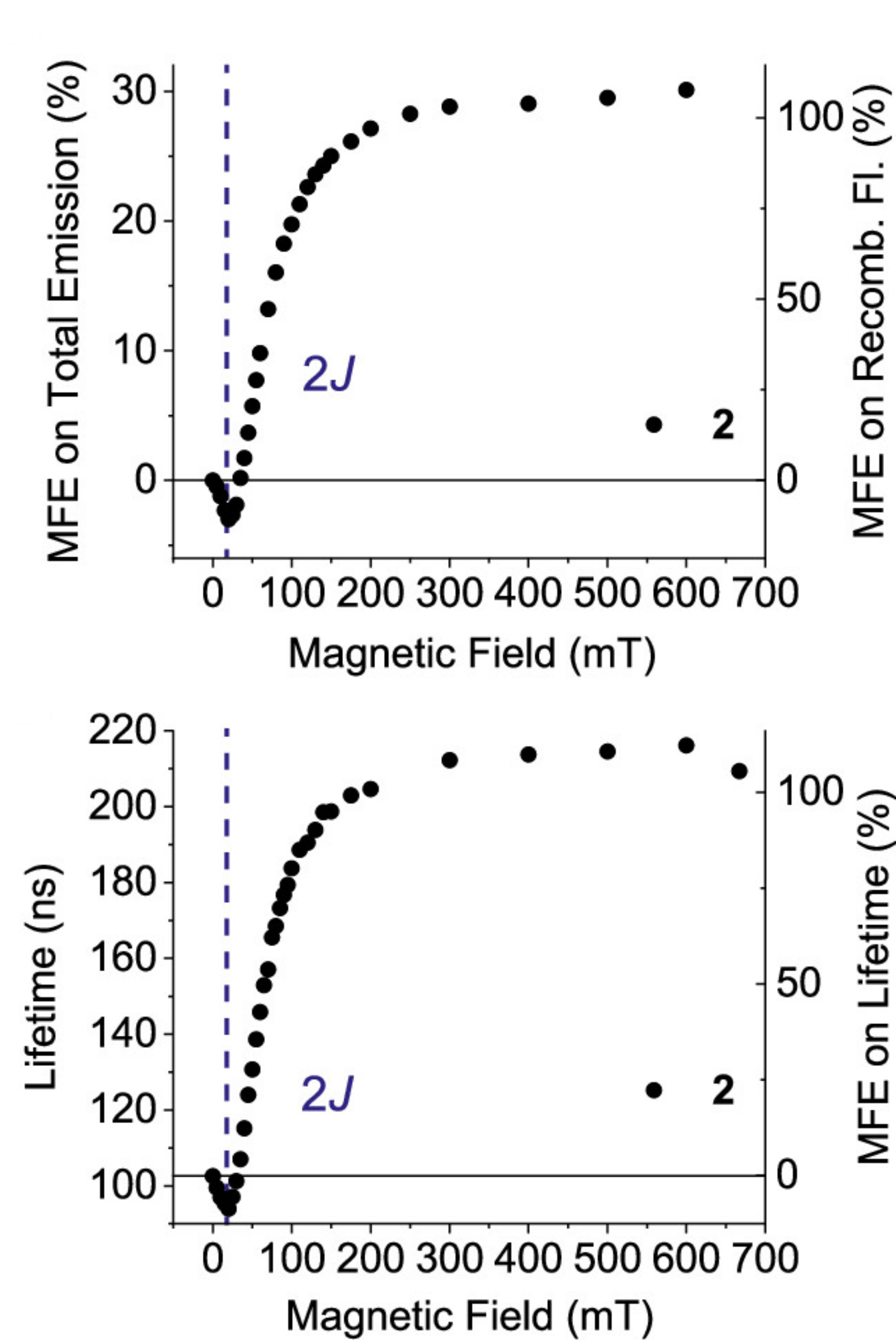}
    \caption{Magnetic modulation of optical properties of a donor-chiral bridge-acceptor molecule. The top plot shows magnetic field effects (MFEs) on total emission (left axis) and recombination fluorescence only (right axis), measured relative to a null field (MFE (\%) $= [\frac{I(B=\mathrm{ON})}{I(B=\mathrm{OFF})}-1] \times 100$, where I is the intensity of emission). The bottom plot shows magnetic modulation of the recombination fluorescence time (left axis) and its MFEs (right axis) (MFE (\%) $= [\frac{\tau(B=\mathrm{ON})}{\tau(B=\mathrm{OFF})}-1] \times 100$, where $\tau$ is the lifetime. Adapted with permission from Lin \textit{et al.}\cite{lin_SCRP_Qubits_2025}, J. Am. Chem. Soc. \textbf{147}, 11062–11071 (2025). Copyright 2025 American Chemical Society.}
    \label{fig:mani}
\end{figure}

Attention has also turned to the potential role of CISS in magnetoreception where, following on from the framework by Luo \& Hore \cite{luo_chiral-induced_2021} introduced in Sec. \ref{sec:electron_transfer}, it has been predicted by Tiwari \& Poonia \cite{tiwari_role_2022, tiwari_quantum_2023} to enhance SCRP sensitivity to magnetic fields, and coherence. The exact mechanism behind these enhancements was unknown, but the most prominent enhancements emerged at strongly asymmetric recombination regimes of the radical pair. Following this, Smith \textit{et al.}\cite{ciss25rp} investigated both singlet-born and triplet-born radical pairs implicated in biological magnetosensing for models of CISS that are selective for polarization or only coherence. They found that CISS-generated spin polarization significantly enhances magnetic field sensitivity by introducing triplet character into the initial state and reinforcing the quantum Zeno effect, which operates at strongly asymmetric recombination rates, while CISS-generated spin coherence was less effective at providing enhancements (Fig.~\ref{fig:ciss_zeno}). This highlights that CISS in itself is not a universal enhancer of sensitivity or coherence, but should be evaluated alongside the quantum Zeno effect in cryptochrome radical pairs, and that the role of generating polarization effects through an incoherent process, could be essential beyond fundamental importance to practical function of CISS in SCRPs. Although the functionality of CISS and SCRPs in biological processes remains an open question, these investigations may help to guide the biomimetic engineering of molecular quantum sensors.     

Recently, Lin \textit{et al.} \cite{lin_SCRP_Qubits_2025} introduced an emerging platform based on donor-chiral bridge-acceptor (D-$\chi$-A) molecules that boast significant magnetic field effects (MFEs) on fluorescence intensity and SCRP lifetimes, even in solution at ambient conditions. These systems build upon previous designs by the Mani group for donor-bridge-acceptor (D-B-A) molecules for optical readout of magnetic-field-sensitive SCRPs at room temperature. While the reported changes for those systems were modest, the D-$\chi$-A  design exhibit significant magnetic field effects on recombination fluorescence, where the chiral bridge provides sufficient electronic couplings to ensure long-lived SCRPs. Although the role of CISS was not explicitly resolved by Lin \textit{et al.}, similar chiral-bridged molecules have been used to investigate and demonstrate the CISS effect, giving credence to the plausibility that CISS could contribute in these systems \cite{eckvahl_holes_2024, eckvahl_direct_2023}. The advantage of molecular synthetic control, was demonstrated through three structural modifications to the donor site via distance extension, torsional locking, and planarization. These changes produced an observed magnetic sensitivity increase of $>200\%$ in emission lifetimes, with up to $30\%$ modulation of total emission intensity (Fig.~\ref{fig:mani}). Furthermore, as these structural changes allow the system rigidity and donor-acceptor distance to be altered, and spin relaxation via singlet-triplet dephasing mechanisms to be modulated, it is plausible that some of the enhancement and control schemes identified in biological SCRPs could be implemented \cite{rwcont25, woods24, prxQC}. D-$\chi$-A SCRPs represent a promising emerging platform of SCRP-based molecular quantum sensors \cite{yu_molecularSensing_2021}, with the potential to extend beyond classical sensing capabilities to address a wide range of applications from biological to condensed matter systems \cite{Aslam2023, Xie_biocompatibleNV_2022, Das_biosensorReview_2024}. Their long-lived spin coherence, room-temperature operation, optical addressability, and structural tunability positions them as strong candidates for developing type-II and type-III quantum sensors that exploit quantum resources such as quantum coherence and entanglement. Moreover, these systems offer a complementary perspective by not only acting as sensors, but also providing testbeds, alongside NV-based quantum sensing to probe fundamental questions about quantum properties of radical pairs and the role of CISS \cite{dasari21, volker2023, deepak_2024}. 
% Quantum phenomena, such as quantum coherence and interference effects (Fano resonances etc.), are generally not expected to survive in noisy ambient conditions, much less in a biological setting. For one, select observations of thermal enhancement of spin-selectivity in CISS-transport could suggest an incoherent hopping mechanism \cite{das_temperature-dependent_2022,sang_temperature_2021, Qian2022-bu}, supporting the treatment of CISS-transport in biomolecules as a semiclassical process. (It should be noted that thermal fluctuations have been argued to weaken CISS instead\cite{alwan_temperature-dependence_2023}.)

% Nevertheless, there is a case for studying possible quantum effects in CISS at ambient temperatures. Quantum interference tends to be robust in molecular junctions, with many proposals for its use\cite{evers_advances_2020}. 

%Another application of CISS is molecular machines. \textcolor{red}{[cite: water splitting Mtangi \textit{et al.}, 2017; spin-induced chiral-selection (origin of life) Ozturk \& Sasselov \textit{et al.}, 2022 / Bloom, Waldeck \& Waldeck, 2022; charge allostery (?) Banerjee-Ghosh \textit{et al.}, 2020; extracellular respiration, Mishra \textit{et al.}, 2019;]}

\section{Concluding Remarks}

CISS has given rise to a rich landscape of experimental findings and promising application since its inception. Though the sheer breadth of CISS studies may obscure the key mechanisms underlying the effect for new entrants, we attempt to lay out a clearer picture of the cumulative efforts into clarifying its origins in an effort to portray a broader point-of-view through which debated issues around CISS may be demystified. At its essence, CISS arises from chiral geometry where finite spin-orbit coupling arises. This generates a spin-texture in reciprocal-space, which gives rise to spin-current when a bias is applied or when charge-current is finite. In an ideal chiral object, this spin-current is entirely non-dissipative and results only in the generation of a magnetic order in the molecule. This is possible because the chiral anomaly breaks conservation of quantized spin-current, contrary to classical-leaning intuitions built on charge-current characteristics.

Naturally, no CISS emerges in equilibrium as no spin-current appears in the absence of external perturbations or applied fields; but it is noteworthy that neither bias alone nor a reversed external magnetic field is sufficient for the \textit{observation} of CISS-transport through biomolecular junctions due to its reliance on magnetocurrent anisotropy as an index of spin-selectivity. This requires both a finite dissipative spin-current and the additional disruption of Onsager-Casimir relations, which may be fulfilled when external dissipation of the system or dephasing is introduced (a core source of investigation from Guo \& Sun \cite{guo_spin-dependent_2014} to Fransson \cite{fransson_chiralOrigin_2025}), or when the setup is simply complicated enough to reasonably discard ideal assumptions. In practice, this is likely to be the case: since bias or charge-current gives rise to a monopole-like spin texture in the chiral object, such a chiral object in quasi-equilibrium is effectively (partially-)ferromagnetised, which contributes to the rich interfacial effects prevalent in CISS experiments. The latter point also accounts for differences in response between CISS in biomolecules (arguably, the `canonical form' of CISS) and CISS in non-doped crystals/metals: even in the absence of bias, wavefunction penetration of a substrate into a chiral bridge can alter electronic properties of the system and give rise to magnetisation-dependent charge-trapping effects in the presence of defect-sites \cite{zhao25}. This dichotomy between dissipation and coherence is featured in other realisations of CISS beyond the realm of molecular junctions: Fay \& Limmer's \cite{fay_origin_2021} SCRP model suggests that spin-orbit coupling gives rise to singlet-triplet coherences which can only then be converted to spin-selectivity by an incoherent electron-transfer step, a finding significant for characterizing enhancements to sensitivity and spin-coherence generation using CISS. Consequently, this may go beyond fundamental interest, to be of essential importance in the development of CISS enabled SCRP-based quantum sensors and may carry across to other CISS based sensing and quantum information applications.

Additionally, it is interesting to see how CISS may be tethered to broader perspectives on chirality: a topological perspective, for instance, shows how gapless Kramers-Weyl fermions naturally appear that are protected by the \textit{lack} of structural inversion-symmetry; and the emergence of entanglement in the ground-state structure thereof make chiral-crystalline materials prime ground for functional quantum effects. A century ago, the first theoretical frameworks for quantum mechanics and electron spin \cite{heisenberg1925, uhlenbeck1925spin} arose in part as a response to the lack of understanding and consensus in regards to anomalies in spectroscopy. So while much remains yet unclear as to CISS's relation to other notions of chirality across different fields; it is an interesting open theoretical question whether some deeper underlying principle, that connects chirality with spin-selectivity, would help us better confront the uneasy gap that currently exists between the formalisms we have and the phenomena we observe across realisations of CISS. Whether this potentially happens to already be utilized by nature is also an open matter at this time.

\section{Acknowledgements}
Y.X.F. acknowledges support from the CNYSP Office at the Nanyang Technological University, Singapore, toward a research visit with the QuBiT Lab at the University of California, Los Angeles. A.K. acknowledges support from an NSF GRFP. F.T.C. and L.D.S. acknowledge support from the Office of Naval Research (ONR Award No. N62909-21-1-2018) and the Engineering and Physical Sciences Research Council (EP/X027376/1). For the purpose of open access, the authors have applied a Creative Commons Attribution (CC BY) license to any Author Accepted Manuscript version arising from this submission.

\section*{Author Declarations}
\subsection*{Conflict of interest}
The authors have no conflicts to disclose.

\section*{Data Availability Statement}
Data sharing is not applicable to this article as no new data were created or analyzed in this study.

%aipnum4-2.bst 2019-01-14 (MD) hand-edited version of apsrev4-1.bst
%Control: key (0)
%Control: author (8) initials jnrlst
%Control: editor formatted (1) identically to author
%Control: production of article title (0) allowed
%Control: page (1) range
%Control: year (1) truncated
%Control: production of eprint (0) enabled
%

\end{document}